\newcommand{\ignore}[1]{}
\documentclass{article}

\usepackage[preprint]{neurips_2026}
\usepackage{graphicx}
\usepackage{tcolorbox}
\usepackage{bm}
\usepackage{amsmath}
\usepackage{multirow}
\usepackage{booktabs}
\usepackage{subcaption}
\usepackage{wrapfig}

\usepackage[utf8]{inputenc} 
\usepackage[T1]{fontenc}    
\usepackage{hyperref}       
\usepackage{url}            
\usepackage{booktabs}       
\usepackage{amsfonts}       
\usepackage{nicefrac}       
\usepackage{microtype}      
\usepackage{xcolor}         

\title{\texttt{EHR-RAGp}: Prototype-Guided Retrieval of Longitudinal Electronic Health Records for Clinical Prediction Models}

\author{%
  Saeed Shurrab$^{1,2}$, Mariam Al-Omari$^{2}$, Dana El Samad$^{2}$, Farah E. Shamout$^{1,2}$
  \\
  New York University$^{1}$ \\
  New York University Abu Dhabi$^{2}$ \\
}

\begin{document}

\maketitle

\begin{abstract}
Electronic Health Records (EHR) contain rich longitudinal patient information and are widely used in predictive modeling applications. However, effectively leveraging historical data remains challenging due to long trajectories, heterogeneous events, temporal irregularity, and the varying relevance of past clinical context. Existing approaches often rely on fixed windows or uniform aggregation, which can obscure clinically important signals. In this work, we introduce \texttt{EHR-RAGp}, a retrieval-based framework that dynamically integrates the most relevant patient history consisting of diverse clinical event types. We propose a prototype-guided retrieval module that acts as an alignment mechanism and estimates the relevance of retrieved historical chunks with respect to a given prediction task, guiding the model towards the most informative context. Across multiple clinical prediction tasks and two benchmark datasets, \texttt{EHR-RAGp} consistently outperforms state-of-the-art EHR-based and transformer-based baselines. Furthermore, \texttt{EHR-RAGp} is model-agnostic, as integrating it with various backbone models yields substantial performance gains. Overall, \texttt{EHR-RAGp} establishes a novel direction for modeling long-range clinical context to improve downstream performance via retrieval.
\end{abstract}

\section{Introduction}

Healthcare systems rely extensively on Electronic Health Record (EHR) databases to keep records of diverse clinical events that occur during patient encounters, such as demographics, diagnoses, vital signs, lab results, and medications \citep{nasarudin2024review}. Recent advances in machine learning have enabled the development of various clinical prediction models for EHR data, where patient trajectories are represented as chronologically ordered sequences of events, in a manner that mimics the structures of sentences in natural language \citep{li2024scoping,ren2025comprehensive, wornow2024context}. However, unlike natural language, clinical events are inherently irregular, and span multiple episodes of care that accumulate over extended periods \citep{zhong2025hybrid}.

\begin{figure*}[t]
    \centering
    \includegraphics[width=1.0\linewidth]{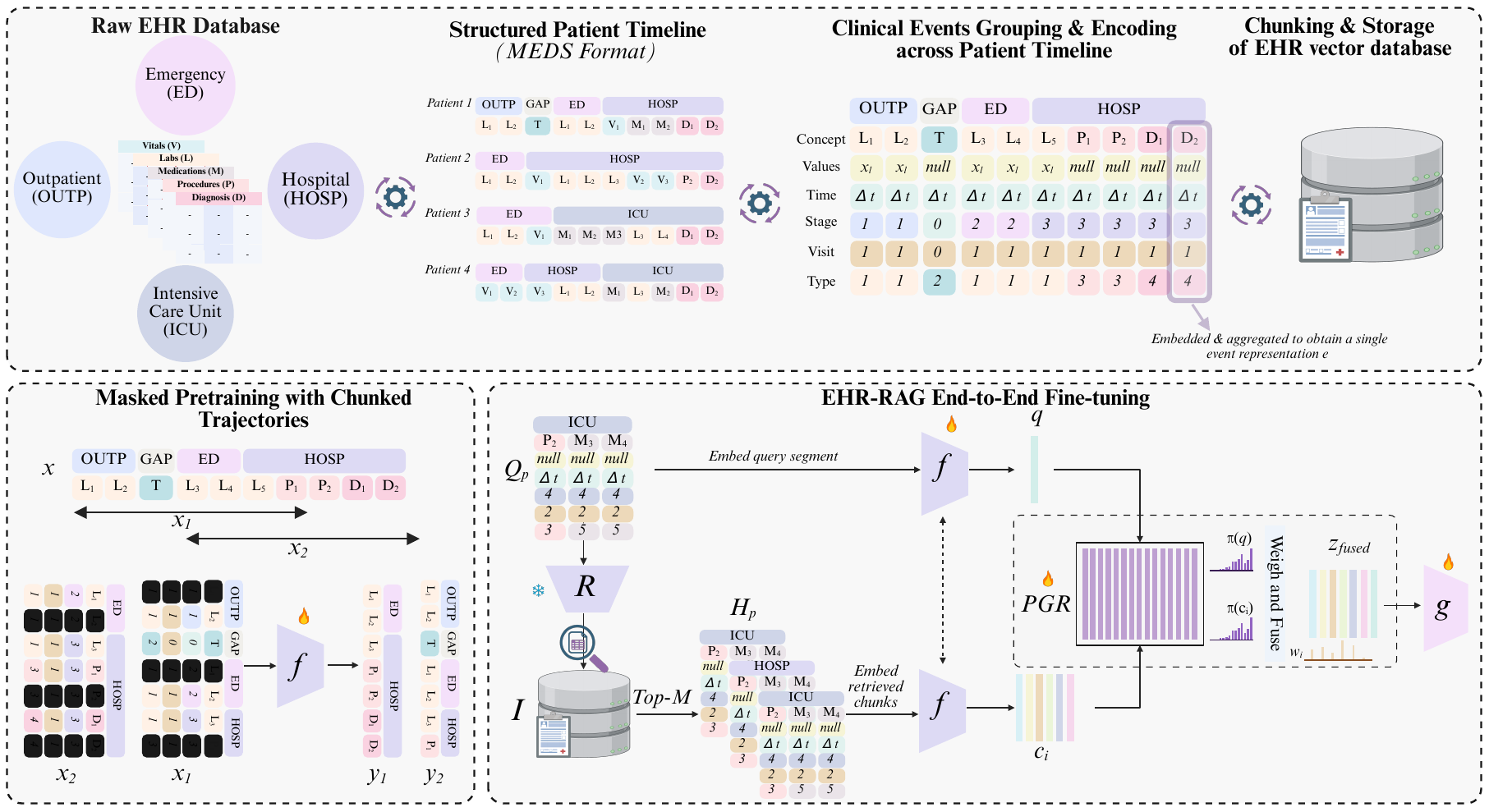}
    \caption{{Overview of the \texttt{EHR-RAGp}, including chunking and indexing of patient trajectories (\textit{top}), masked pre-training of the backbone encoder (\textit{bottom left}), and end-to-end fine-tuning of the final architecture (\textit{bottom right}).}} 
    \vspace{-6mm}
    \label{fig:main-figure}
\end{figure*}
The current state of EHR modeling is shaped by specific design assumptions, including: (i) information present only within a predefined time window for a specific patient visit is relevant to predict the outcome of interest \citep{wornow2023ehrshot, elsharief2025medmod}; and (ii) a preselected subset of events or features, such as labs, vitals, diagnoses, or combinations of them, is sufficient to predict patient outcomes \citep{ odgaard2024core, hur2023genhpf, yang2023transformehr}. Although these assumptions provide practical abstractions, they may restrict the model’s ability to capture relevant information beyond predefined temporal windows or feature sets. Beyond these design assumptions, computational constraints and the difficulty of representing long-range, irregular patient trajectories present additional barriers to comprehensive EHR modeling \citep{wornow2024context}, which may also lead to the omission of clinically important historical events. From a clinical perspective, the patient’s state is almost always shaped by their accumulated longitudinal history \citep{braitman1996predicting, duan2019clinical}. This includes slow-moving chronic patterns and cross-encounter interactions that clinicians routinely examine when making decisions. Such a central gap highlights a fundamental limitation in current EHR modeling paradigms, namely, the inability to integrate and reason over a patient’s extended history within a unified representation learning framework.

Filling this gap requires a new outlook on EHR representation learning. Rather than limiting the patient’s predictive context to a predefined window, a model must be equipped to access, retrieve, and integrate information present in the patient’s complete trajectory. The central assumption motivating this work is that retrieval provides a scalable mechanism for incorporating long-range historical context without forcing the model to encode the entire trajectory within its fixed internal state. Given the remarkable success of Retrieval-Augmented Generation \citep{lewis2020retrieval} (RAG) techniques in Natural Language Processing (NLP) and their ability to leverage external knowledge beyond the training data of the base Large Language Model (LLM) \citep{yu2024evaluation,zhao2024retrieval,zheng2025retrieval}, the same concept can be applied for retrieving and integrating historical EHR data. In other words, by considering the patient’s historical events as retrievable units of information, an EHR model can dynamically condition on what matters the most for the patient’s current state. This perspective shifts longitudinal modeling from a purely sequential problem into a retrieval-augmented one. Hence, in this work, we propose {\texttt{EHR-RAGp}}, a retrieval-based EHR modeling framework designed to reason over extended context of a patient’s longitudinal EHR history. The main contributions of the work are summarized as follows:
\vspace{-2mm}
\begin{enumerate}
    \item We establish modeling of long EHR records as a retrieval problem, by constructing \textit{a vector database of longitudinal EHR patient trajectories} using multi-granular chunking strategies, including event-based, time-based, visit-level, and care-stage, enabling flexible representation of heterogeneous events, temporal structure, and care progression.
    \item We present \texttt{EHR-RAGp}, a \textit{prototype-guided retrieval mechanism} that acts as an alignment operator, estimating the relevance of retrieved patient chunks with respect to the downstream task, guiding the model to integrate the most informative retrieved clinical context.
    \item Extensive evaluation across \textit{multiple clinical prediction tasks and two datasets} demonstrates that \texttt{EHR-RAGp} consistently improves performance over multiple clinical and general baselines. 
    \item We demonstrate the effectiveness of \texttt{EHR-RAGp} as a \textit{model-agnostic framework}, by experimenting with multiple EHR backbone models, improving their performance beyond fixed context. 
    \item Existing EHR models vary in their selection and preprocessing of input features.
    By training all baselines using a \textit{unified feature set and training pipeline}, we establish new benchmark results on four clinical prediction tasks and make our code publicly available at:
     \href{https://github.com/nyuad-cai/EHR-RAGp}{https://github.com/nyuad-cai/EHR-RAGp}.
\end{enumerate}

\section{Related Work}

\textbf{EHR Modeling.}  Traditional EHR modeling methods process patient timelines as sequences of discrete events collected over time. Some models primarily consider the occurrence of certain events, while others process the occurrence of the clinical event and its measured value. Early techniques utilized recurrent models such as Long-Short Term Memory (LSTM) \citep{hochreiter1997long} or Gated Recurrent Unit (GRU) networks \citep{cho2014properties} to capture long-term dependencies \citep{che2018recurrent,harutyunyan2019multitask, duan2019clinical,hayat2022medfuse}.

The emergence of transformer-based LLMs \citep{devlin2019bert, radford2019language} paved the way for advancing EHR modeling. For example, encoder-based models such as BEHRT \citep{li2020behrt}, Hi-BEHRT \citep{li2022hi}, CEHR-BERT \citep{pang2021cehr}, and Med-BERT \citep{rasmy2021med} introduced large-scale pretraining over longitudinal sequences of medical events to enhance the contextualization of patient trajectories across visits. Other decoder-based transformer models such as ETHOS \citep{renc2025foundation}, CEHR-GPT \citep{pang2024cehr}, Curiosity \citep{waxler2025generative}, and EHR2Path \citep{pellegrini2025ehrs} focused on the development of generative EHR foundation models for clinical prediction tasks. The common link between all these models is that they mimic standard language models by learning contextualized medical code vocabularies as tokens.

Another category of EHR foundation models shifts away from code-based modeling towards text-based modeling by treating each medical code as a textual description. Hence, the patient timeline constitutes a set of textual descriptions that can be processed directly by the LLM \citep{hur2022unihpf, hur2023genhpf, lee2025clinical}. Despite the remarkable progress, these models continue to inherit core limitations from their language-model roots. They operate on truncated encounter windows, compress long and irregular patient histories into fixed-length sequences, and rely on restricted types of clinical events. While some methods have explored the use of longer context windows to accommodate extended trajectories \citep{waxler2025generative,hur2022unihpf, hur2023genhpf}, their vocabularies are heavily tokenized for medical codes and numerical values, or textualized and dependent on the base model's reasoning capabilities. As a result, substantial portions of a patient’s longitudinal trajectory remain inaccessible in downstream prediction tasks. This emphasizes the need for scalable frameworks that are capable of integrating and reasoning over complete EHR trajectories.

\textbf{Retrieval-Augmented Modeling.} RAG has emerged as a powerful technique in NLP for extending large language models capabilities by allowing them to condition on external, non-parametric knowledge sources. The original RAG framework \citep{lewis2020retrieval} demonstrated how combining dense retrieval with generative transformers improves factual correctness and reduces hallucinations. The concept was later extended by models such as RETRO \citep{borgeaud2022improving}, FLARE \citep{jiang2023active}, and REPLUG \citep{shi2024replug}. These architectures share a common principle: retrieval allows the model to incorporate long-range or domain-specific information without encoding the entire knowledge base within its internal parameters. The success of the retrieval paradigm has also influenced healthcare, particularly in clinical question answering \citep{saba2024question, sohn2025rationale}, evidence grounding \citep{lewis2025grounding, jia2025agentic}, biomedical knowledge integration \citep{feng2025retrieval, feng2025ontologyrag}, phenotyping and cohort identification \citep{ziletti2025generating}, and many other applications related to medical textual data \citep{abo2025survey, amugongo2025retrieval}.

 In the context of EHR modeling, retrieval has been  explored in work such as EMERGE \citep{zhu2024emerge}, REALM \citep{zhu2024realm}, and RAM-EHR \citep{xu2024ram}. The goal of these studies is to incorporate knowledge from external related sources into their modeling pipelines to enhance predictive accuracy. Hence, the retrieval objective of these studies is not aligned with our scope, \textit{which specifically pertains to retrieval over the patient's history.} Furthermore, these models operate on multimodal data beyond EHR events only, like notes. To the best of our knowledge,  REMED \citep{kim2023general} is most closely related to our work. Its core idea is to retrieve the most relevant subset of medical codes for a given outcome prediction task, to overcome manual selection of clinical events or features. Overall, recent advancements highlight the potential of retrieval-augmented modeling for EHR, but they also reveal a clear unmet need for mechanisms that can dynamically access and integrate a patient’s entire structured history.

\section{Methodology}
\label{sec:method}

\vspace{-1mm}
\subsection{Overview}
\vspace{-1mm}
\texttt{EHR-RAGp} is a retrieval-augmented model for reasoning over longitudinal EHR trajectories (Figure~\ref{fig:main-figure}). The framework encodes structured clinical events, organizes them into retrievable units, retrieves relevant historical segments through prototype-guided alignment, and integrates them with a task-specific query for downstream task. Here, we describe the core architectural components.

\subsection{Preliminaries}

Let  $\mathcal{D}=\{\mathcal{H}_p \}_{p=1}^{\mathcal{P}}$ represent a large-scale EHR dataset consisting of patient-level longitudinal records for a cohort of $\mathcal{P}$ patients. For a given patient $p$, the longitudinal medical history is represented as:
\begin{equation*}
    \mathcal{H}_p = \{\mathcal{V}_p^{(1)}, \mathcal{V}_p^{(2)}, \ldots, \mathcal{V}_p^{(N_p)}\},
\end{equation*}
where $N_p$  denotes the total number of visits for the respective patient (see Figure~\ref{fig:sample-seq}).  The $n$-th visit  $\mathcal{V}_p^{(n)}$ consists of a set of clinical events:
\begin{equation*}
    \mathcal{V}_p^{(n)} = \{ (c_{nk}, v_{nk}, t_{nk}) \}_{k=1}^{K_n},
\end{equation*}
where $c_{nk}$ denotes a clinical event such as a diagnosis, procedure, laboratory test, vital sign, or medication administration, $v_{nk}$ is an optional associated value such as a lab result, medication dosage, or vital sign measurement, and $t_{nk}$ is the exact timestamp at which the event occurred.  Events within a visit may vary in type, granularity, and frequency, reflecting the inherently irregular and heterogeneous nature of structured EHR data.

For a given index visit $\mathcal{V}_p^{(\tau)}$, the goal is to predict a target patient outcome with true label $y_p^{(\tau)}$ by retrieving the most relevant historical information. We define a \textbf{query segment} $\mathcal{Q}_p^{(\tau)} \subseteq \mathcal{V}_p^{(\tau)}$ as the portion of the current visit used as the model’s immediate input. Since the full visit may exceed the model’s maximum input length, the query is restricted to a fixed-size segment within its context window. All preceding events, whether from the same visit or earlier visits, form the \textbf{historical context}:
\begin{equation*}
    \mathcal{H}_p^{(<\tau)} = \{ \mathcal{V}_p^{(1)}, \ldots, \mathcal{V}_p^{(\tau-1)}, \ \mathcal{V}_p^{(\tau<q)} \},
\end{equation*}
where $\mathcal{V}_p^{(\tau<q)}$ denotes the prefix of the last visit preceding the query (see Figure \ref{fig:query-history}). Hence, the model aims to learn a mapping function:
\begin{equation*}
    \gamma :\big(\mathcal{Q}_p^{(\tau)}, \mathcal{H}_p^{(<\tau)}\big) \rightarrow \hat{y}_p^{(\tau)}.
\end{equation*}
A naive formulation of this problem would require the model to encode the full patient history in its fixed internal state, which is a significant challenge due to computational constraints. Hence, the goal of our proposed framework is to enable the model to retrieve and integrate the most relevant information from the patient's history $\mathcal{H}_p^{(<\tau)}$ based on the query segment $\mathcal{Q}_p^{(\tau)}$.

\vspace{-2mm}
\subsection{Chunking}
After representing individual events, \texttt{EHR-RAGp} partitions the patient timeline into coherent chunks that serve as retrievable units. We propose four chunking methods, namely,  \textit{event-based}, \textit{time-based}, \textit{visit-level}, and \textit{care-stage} chunking (Figure \ref{fig:chunking-methods}). Each chunking method captures a different aspect of longitudinal patient structure, enabling retrieval over heterogeneous temporal granularities and clinical contexts. Event-based chunking groups a fixed number of consecutive events, time-based chunking partitions the timeline into fixed temporal windows, visit-level chunking organizes events according to hospital admissions, and care-stage chunking separates events based on transitions between clinical care settings, such as outpatient, emergency department, hospital ward, and ICU. 

\vspace{-2mm}
 
\subsection{Prototype-Guided Retrieval}
\label{sec:proto}
Given a query segment $\mathcal{Q}_p^{(\tau)}$, \texttt{EHR-RAGp} retrieves relevant historical segments from the patient’s longitudinal history $\mathcal{H}_p^{(<\tau)}$ using a two-stage procedure: (1) coarse-grained semantic retrieval to identify candidate chunks, followed by (2) prototype-guided alignment that estimates the relevance of each retrieved chunk via agreement in the learned prototype assignment space, to enable weighted integration of context for the downstream task.

\textbf{Candidate Retrieval.}
The query segment $\mathcal{Q}_p^{(\tau)}$ is first encoded using a frozen pretrained  retriever $R$, producing a fixed-dimensional query representation:
\begin{equation*}
q' = R(\mathcal{Q}_p^{(\tau)}) \in \mathbb{R}^{d}.
\end{equation*}

 The retriever, $R$, is pretrained using masked language modeling. Further details are provided in Appendix~\ref{sec: exp-setup}. Using the same retriever $R$, each historical chunk $h_i \in \mathcal{H}_p^{(<\tau)}$ is embedded as
\begin{equation*}
c'_i = R(h_i) \in \mathbb{R}^{d}.
\end{equation*}
Candidate retrieval is then performed by computing \textbf{cosine} similarity between the query embedding $q'$ and historical chunk embeddings $\{c'_i\}$ stored in precomputed similarity index $\mathcal{I}$. We then retrieve the top-$M$ most similar chunks:
\begin{equation*}
\{h_1, h_2, \ldots, h_M\} \subset \mathcal{H}_p^{(<\tau)}.
\end{equation*}

\textbf{Query and Candidate Encoding.} Upon retrieval, the query segment $\mathcal{Q}_p^{(\tau)}$ and the top-$M$ chunks  $\{h_1, h_2, \ldots, h_M\}$ are passed to a shared backbone encoder $f$ to obtain $q$ and $\{c_i\}$, respectively.

\begin{equation*}
q = f(\mathcal{Q}_p^{(\tau)}) \in \mathbb{R}^{d},
\quad\quad
c_i = f(h_i) \in \mathbb{R}^{d}.
\end{equation*}
Using the same encoder $f$ for queries and history ensures that both are represented in a common space, enabling consistent comparison and alignment between the query and retrieved context.

\textbf{Prototype-Guided Alignment.}
\texttt{EHR-RAGp} maintains a learnable set of $L$ prototypes, which serve as latent clinical centroids that organize a patient’s longitudinal history into distinct semantic modes:
\begin{equation*}
P = \{p_1, p_2, \ldots, p_L\}, \qquad p_l \in \mathbb{R}^{d},
\end{equation*}
These prototypes serve as latent anchors that organize the representation space into distinct modes, enabling structured comparison between query and historical chunks. Rather than being fixed, they are learned end-to-end, allowing the model to capture recurring patterns in clinical trajectories and facilitate alignment between the query and retrieved context with respect to the downstream task.

For a chunk embedding $x \in \mathbb{R}^{d}$,  the model computes a probability distribution over the $L$ prototypes:
\begin{equation*}
\pi(x) = \operatorname{softmax}\left( \frac{x P^\top}{T_\pi} \right) \in \mathbb{R}^{L},
\end{equation*}
where $x$  corresponds to either to the \textbf{query} $q$ or a \textbf{candidate chunk} $c_i$, and $T_\pi$ is a temperature parameter that controls the sharpness of the distribution \citep{assran2022masked}. In practice, we use different temperatures for the query and history branches, denoted by $T_q$ and $T_h$, respectively, with $T_q < T_h$. This asymmetric design reflects the roles of the two branches: the query must identify a focused subset of relevant patterns for the prediction task, while historical chunks represent diverse clinical contexts that may only partially match the query. Accordingly, we obtain:

\begin{equation*}
\pi_q = \pi(q), \qquad \pi_i = \pi(c_i), \; i = 1, \ldots, M.
\end{equation*}
$\pi_q$ represents the prototype assignment distribution of the query, while each $\pi_i$ represents the assignment of candidate chunk $c_i$ to the set of learned prototypes \textit{L}. To measure how well a candidate history chunk aligns with the query in the prototype space, \texttt{EHR-RAGp} measures the agreement between their prototype distributions using a cross-entropy-based alignment score:
\begin{equation*}
\alpha_i = - \sum_{l=1}^{L} \pi_q^{(l)} \log \pi_i^{(l)},
\end{equation*}

Where $\alpha \in [0, \infty) $.
This formulation measures how well the candidate distribution $\pi_i$ explains the query distribution $\pi_q$, treating the query as a reference. Lower negative cross-entropy values indicate stronger alignment, meaning that the candidate assigns high probability mass to the prototypes emphasized by the query, whereas higher values indicate weaker alignment. In addition, this distributional alignment enables smooth and differentiable matching in the prototype space without requiring hard assignment to a single prototype.

\textbf{Prototype-Guided Weighting.} 
Given the prototype alignment scores $\{\alpha_i\}_{i=1}^M$, \texttt{EHR-RAGp} converts them into normalized importance weights using a temperature-controlled softmax:
\begin{equation*}
w_i = \operatorname{softmax}\!\left(\frac{-\alpha_i}{T_s}\right),
\end{equation*}
where $T_s$ is a temperature parameter that controls the smoothness of the weighting distribution.

\texttt{EHR-RAGp} adopts a fully differentiable soft weighting mechanism, where all retrieved chunks contribute to the final prediction with varying degrees of importance. This design allows the model to preserve distributed clinical evidence across multiple chunks, which is particularly important in EHR data where relevant signals are often weak and spread over time. The resulting weights $\{w_i\}$ are used to modulate the contribution of each retrieved chunk during the subsequent fusion stage.

\textbf{Retrieval-Augmented Fusion.}
The query representation $q$ and the retrieved historical chunks $\mathcal{R}_p^{(<\tau)} = \{c_1, \ldots, c_n\}$ are combined through a prototype-guided weighting mechanism. Each retrieved chunk is modulated by its corresponding importance weight $w_i$:
\begin{equation*}
    z_{\text{fused}} = \big[q,\; w_1 c_1,\; w_2 c_2,\; \ldots,\; w_n c_n \big].
\end{equation*}

This sequence is then processed by a transformer encoder $g(\cdot)$  followed by a a linear classifier $h(\cdot)$  to yield the final prediction:
\begin{equation*}
    \hat{y}_p^{(\tau)} = h\left(g\!\left(z_{\text{fused}}\right)\right).
\end{equation*}

\textbf{Training Objective.} \texttt{EHR-RAGp} is trained end-to-end using a supervised binary cross-entropy loss associated with the target prediction task, while keeping the retrieval encoder, $R$, frozen and updating the backbone encoder, $f$, the prototype parameters, and the fusion module. The overall objective combines the task loss with a prototype usage regularization term that encourages balanced utilization of the prototype space across both query and history representations:
\begin{equation*}
\mathcal{L} = l\big(y_p^{(\tau)}, \hat{y}_p^{(\tau)}\big)
- \lambda_{\text{u}} \left( \mathcal{H}(\bar{\pi}_q) + \mathcal{H}(\bar{\pi}_h)\right),
\end{equation*}
where $\bar{\pi}_q = \frac{1}{B} \sum_{b=1}^{B} \pi_q^{(b)}$ and $\bar{\pi}_h = \frac{1}{\sum_b K_b} \sum_{b,k} \pi_{h_k}^{(b)}$ denote the average prototype assignment distributions over a batch for the query and historical chunks, respectively, $\mathcal{H}(\cdot)$ is the entropy function, and $\lambda_{\text{u}}>0$ determines the  strength of the regularization term. This regularization encourages the model to avoid degenerate solutions in which only a small subset of prototypes are used.

\section{Experiments}
\textbf{Datasets \& Tasks.} We evaluate our proposed framework on multiple tasks using the large-scale real-world MIMIC-IV dataset (V3.1) \citep{johnson2020mimic}, and EHRSHOT \citep{wornow2023ehrshot} dataset. MIMIC-IV consists of heterogeneous EHR data gathered from patients admitted to the Beth Israel Deaconess Medical Center between 2008-2022. It includes clinical data collected from 364K patients associated with 546K hospital admissions and 94K ICU stays \citep{johnson2023mimic}. EHRSHOT is a longitudinal dataset for performance benchmarking  of foundation models for clinical prediction tasks. EHRSHOT covers data from 6,739 patients, 921K visits, and 15 tasks.  


For the MIMIC-IV dataset, we assess \texttt{EHR-RAGp} on four risk prediction tasks, including: in-hospital mortality, ICU-readmission within 30 days, long length of stay (7 days or more), and 1-year mortality post-discharge. The selected tasks span both short-term and long-term clinical outcomes and reflect common benchmarks in EHR-based modeling. For EHRSHOT dataset, we consider the assignment of new diagnosis task family, including: Acute MI (MI), Lupus (LP), Hyperlipidemia (HL),  Hypertension (HT), Celiac (CEL), and Pancreatic Cancer (PC). We also consider operational outcome task family including: ICU admission (ICU), long length of stay $>$7 days (LLOS),  and 30-day Readmission (RA). Together, these tasks assess the model’s ability to capture acute risk, longitudinal disease progression, and healthcare utilization by leveraging structured patient histories. Detailed information pertaining to the tasks and data preprocessing are in Appendix \ref{sec:tasks}.

\begin{table*}[t!]
\centering
\caption{Performance of \texttt{EHR-RAGp} and baselines on the MIMIC-IV dataset. Best results are marked in \textbf{bold}. Second-best results are \underline{underlined} (ties allowed). $^\dagger$denotes p$<$ 0.05 compared to second-best.}
\small
\setlength{\tabcolsep}{5pt}
\renewcommand{\arraystretch}{1.05}
\resizebox{0.9\linewidth}{!}{
\begin{tabular}{lcccccccc}
\toprule
\multirow{1}{*}
& \multicolumn{2}{c}{\textbf{ICU-Readmit 30d}}
& \multicolumn{2}{c}{\textbf{In-hospital Mortality}}
& \multicolumn{2}{c}{\textbf{Long LOS 7d}}
& \multicolumn{2}{c}{\textbf{1YR Mortality}} \\
\cmidrule(lr){2-3}
\cmidrule(lr){4-5}
\cmidrule(lr){6-7}
\cmidrule(lr){8-9}

{\textbf{Model}}
& \textbf{AUROC} & \textbf{AUPRC}
& \textbf{AUROC} & \textbf{AUPRC}
& \textbf{AUROC} & \textbf{AUPRC}
& \textbf{AUROC} & \textbf{AUPRC} \\
\midrule

DescEmb I \ignore{\citep{hur2022unifying}}
& 0.661 & 0.078
& 0.904 & 0.618
& 0.822 & 0.405
& 0.754 & 0.376 \\

DescEmb II \ignore{\citep{hur2022unifying}}
& 0.606 & 0.071
& 0.920 & 0.648
& 0.838 & 0.441
& 0.805 & 0.471 \\

GenHPF \ignore{\citep{hur2023genhpf}}
& 0.697 & \underline{0.128}
& 0.915 & 0.641
& 0.843 & 0.456
& 0.781 & 0.417 \\

Med-BERT \ignore{\citep{rasmy2021med}}
& 0.664 & 0.081
& 0.893 & 0.579
& 0.813 & 0.412
& 0.729 & 0.338 \\

REMed \ignore{\citep{kim2023general}}
& 0.561 & 0.060
& 0.865 & 0.501
& 0.786 & 0.348
& 0.680 & 0.280 \\

CEHR-BERT \ignore{\citep{pang2021cehr}}
& 0.604 & 0.061
& \underline{0.929} & \underline{0.676}
& 0.840 & 0.435
& \underline{0.806} & \underline{0.473} \\

BEHRT \ignore{\citep{li2020behrt}}
& 0.668 & 0.077
& 0.902 & 0.609
& 0.824 & 0.444
& 0.767 & 0.382 \\

Hi-BEHRT \ignore{\citep{li2022hi}}
& 0.620 & 0.069
& 0.886 & 0.542
& 0.798 & 0.392
& 0.703 & 0.298 \\

EHRMamba \ignore{\citep{fallahpour2024ehrmamba}}
& 0.665 & 0.082
& \underline{0.929} & 0.650
& 0.829 & 0.424
& 0.796 & 0.429 \\

ModernBERT \ignore{\citep{warner2025smarter}}
& 0.604 & 0.068
& 0.920 & 0.640
& 0.827 & 0.440
& 0.765 & 0.381 \\

LongFormer \ignore{\citep{beltagy2020longformer}}
& 0.645 & 0.095
& 0.910 & 0.577
& 0.821 & 0.423
& 0.760 & 0.377 \\

BigBird \ignore{\citep{zaheer2020big}}
& 0.719 & 0.118
& 0.913 & 0.643
& 0.830 & 0.444
& 0.787 & 0.424 \\

RoFormer \ignore{\citep{su2024roformer}}
& 0.699 & 0.114
& 0.927 & 0.672
& \underline{0.846} & 0.461
& \underline{0.806} & 0.460 \\

RoBERTa \ignore{\citep{liu2019roberta}}
& \underline{0.721} & 0.125
& 0.925 & 0.671
& 0.841 & \underline{0.465}
& 0.805 & 0.466 \\
\midrule

\texttt{EHR-RAGp} (Ours)
& \textbf{0.750}$^\dagger$ & \textbf{0.141}
& \textbf{0.950}$^\dagger$ & \textbf{0.736}$^\dagger$
& \textbf{0.867}$^\dagger$ & \textbf{0.509}$^\dagger$
& \textbf{0.853}$^\dagger$ & \textbf{0.564}$^\dagger$ \\

\bottomrule

\vspace{-3mm}
\end{tabular}}
\label{tab:mimic_results}
\end{table*}

\begin{table*}[t]
\centering
\caption{Performance of \texttt{EHR-RAGp} and baselines on the EHRSHOT dataset in terms of AUROC. Best results are marked in \textbf{bold}. Second-best results are \underline{underlined} (ties allowed). $\Delta$ represents performance difference compared to the base backbone CLMBR.
}
\label{tab:retrieval_long_context}
\scriptsize
\setlength{\tabcolsep}{3.5pt}
\renewcommand{\arraystretch}{1.1}

\resizebox{0.87\textwidth}{!}{
\begin{tabular}{lccccccccc}
\toprule
\multirow{2}{*} &
\multicolumn{3}{c}{\textbf{Operational tasks}} &
\multicolumn{6}{c}{\textbf{Diagnostic tasks}} \\
\cmidrule(lr){2-4}
\cmidrule(lr){5-10}

{\textbf{Model}}
& \textbf{ICU}
& \textbf{LLOS}
& \textbf{RA}
& \textbf{MI}
& \textbf{LP}
& \textbf{HL}
& \textbf{HT}
& \textbf{CEL}
& \textbf{PC} \\
\midrule

LLaMA
& 0.845
& 0.810
& \textbf{0.823}
& 0.744
& 0.724
& 0.691
& \underline{0.722}
& \underline{0.783}
& 0.805 \\

GPT-3
& \underline{0.859}
& 0.813
& 0.814
& 0.735
& 0.739
& 0.687
& \underline{0.722}
& 0.706
& 0.840 \\

Hyena
& 0.738
& 0.766
& 0.762
& 0.629
& 0.602
& 0.659
& 0.647
& 0.733
& 0.740 \\

Mamba
& 0.855
& \underline{0.827}
& \underline{0.816}
& \underline{0.745}
& \underline{0.784}
& \underline{0.698}
& 0.721
& 0.751
& \underline{0.866} \\

CLMBR
& 0.848
& 0.814
& 0.810
& 0.729
& 0.747
& 0.675
& 0.718
& 0.557
& 0.813 \\

\midrule

\texttt{EHR-RAGp}
& \textbf{0.881}
& \textbf{0.845}
& \underline{0.816}
& \textbf{0.761}
& \textbf{0.863}
& \textbf{0.721}
& \textbf{0.749}
& \textbf{0.785}
& \textbf{0.873} \\

\midrule

$\Delta$
& $+$3.89\%
& $+$3.80\%
& $+$0.74\%
& $+$4.39\%
& $+$15.53\%
& $+$6.81\%
& $+$4.31\%
& $+$41.0\%
& $+$7.38\% \\

\bottomrule
\end{tabular}
}
\label{tab:ehrshot_results}
\vspace{-2mm}
\end{table*}

\textbf{Implementation Details.} We use RoBERTa \textit{base} \citep{liu2019roberta} as the backbone encoder for \texttt{EHR-RAGp}, for the MIMIC-IV experiments. In the first stage, we pretrain the encoder using an MLM objective. For fairness of comparison, all code-based baselines in our evaluation are pretrained using the same objective and data splits. The pretrained \texttt{EHR-RAGp} encoder is subsequently reused as the embedding model for constructing and querying the similarity index. In the second stage, we fine-tune the full \texttt{EHR-RAGp} model end-to-end for the downstream prediction tasks in a supervised setting. For the EHRSHOT dataset, we integrate CLMBR \citep{wornow2023ehrshot}, a pretrained model, published with the dataset, in the \texttt{EHR-RAGp} pipeline as a backbone. We conduct extensive hyperparameter tuning using Bayesian optimization for all baselines and \texttt{EHR-RAGp} models. Detailed optimization settings and training schedules are provided in Appendix~\ref{sec: exp-setup}. We evaluate MIMIC-IV experiments using the Area Under the Operator Receiver Curve (AUROC) and the Area Under the Precession-Recall (AUPRC). We also report statistical significance based on paired bootstrap testing. We use AUROC only to evaluate EHRSHOT experiments, based on existing results in previous work \citep{wornow2024context}.


\textbf{Baselines.} Our main goal is to compare fixed context length EHR models to our proposed retrieval approach. Hence, we consider a specific set of baselines spanning clinical EHR models, and general transformer models. We also examine how LLMs perform in a zero-shot fashion using fixed context. Clinical baselines include existing EHR-specific models such as DescEmb \citep{hur2022unifying}, GenHPF \citep{hur2023genhpf}, Med-BERT \citep{rasmy2021med}, CEHR-BERT \citep{pang2021cehr}, BEHRT \citep{li2020behrt},  Hi-BEHRT \citep{li2022hi}, EHRMamba \citep{fallahpour2024ehrmamba}, and REMed \citep{kim2023general}. We denote DescEmb I as the CLS-FT variant, and DescEmb II as the BERT-FT variant. Since these models were originally trained on different subsets of clinical features, we adapt all baselines using a shared vocabulary and EHR embedding layer to ensure unified input representations and fair comparison. general transformer models include RoBERTa \citep{liu2019roberta}, Longformer \citep{beltagy2020longformer}, BigBird \citep{zaheer2020big}, RoFormer \citep{su2024roformer}, and ModernBERT \citep{warner2025smarter}. Finally, we evaluate both general-purpose (3 models) 
and medically adapted LLMs (3 models)
For EHRSHOT experiments. We also compare EHRSHOT results with long context baselines including LLaMA \citep{grattafiori2024llama}, GPT-3 \citep{brown2020language}, Hyena \citep{poli2023hyena}, Mamba \citep{gu2024mamba} with contexts length of $4K$, $4K$, $16K$, $16K$, respectively.  Additional details are provided in Appendix \ref{sec:baselines}.


 \begin{table*}[!t]
\centering
\caption{Performance gains of existing EHR baselines with \texttt{EHR-RAGp}. Best results are shown in \textbf{bold}. $^\dagger$denotes p$<$ 0.05 compared to the base model.}
\resizebox{0.9\linewidth}{!}{
\begin{tabular}{lccccccccc}
\toprule
\multirow{2}{*}{\textbf{Model}}
& \multirow{2}{*}{{\texttt{EHR-RAGp}}}
& \multicolumn{2}{c}{\textbf{ICU-Readmit 30d}}
& \multicolumn{2}{c}{\textbf{In-hospital Mortality}}
& \multicolumn{2}{c}{\textbf{Long LOS 7d}}
& \multicolumn{2}{c}{\textbf{1YR Mortality}} \\
\cmidrule(lr){3-4}
\cmidrule(lr){5-6}
\cmidrule(lr){7-8}
\cmidrule(lr){9-10} 
& & \textbf{AUROC} & \textbf{AUPRC}
& \textbf{AUROC} & \textbf{AUPRC}
& \textbf{AUROC} & \textbf{AUPRC}
& \textbf{AUROC} & \textbf{AUPRC} \\
\midrule

\multirow{2}{*}{Med-BERT}
& $\times$
& 0.664 & 0.081
& 0.893 & 0.579
& 0.813 & 0.412
& 0.729 & 0.338 \\
& $\surd$
& \textbf{0.715}$^\dagger$ & \textbf{0.117}$^\dagger$
& \textbf{0.938}$^\dagger$ & \textbf{0.702}$^\dagger$
& \textbf{0.843}$^\dagger$ & \textbf{0.455}$^\dagger$
& \textbf{0.823}$^\dagger$ & \textbf{0.476}$^\dagger$ \\

\midrule

\multirow{2}{*}{CEHR-BERT}
& $\times$
& 0.604 & 0.061
& 0.929 & 0.676
& 0.840 & 0.435
& 0.806 & 0.473 \\
& $\surd$
& \textbf{0.698}$^\dagger$ & \textbf{0.102}$^\dagger$
& \textbf{0.941}$^\dagger$ & \textbf{0.715}$^\dagger$
& \textbf{0.848}$^\dagger$ & \textbf{0.452}
& \textbf{0.821}$^\dagger$ & \textbf{0.495}$^\dagger$ \\

\midrule

\multirow{2}{*}{BEHRT}
& $\times$
& 0.668 & 0.077
& 0.902 & 0.609
& 0.824 & 0.444
& 0.767 & 0.382 \\
& $\surd$
& \textbf{0.723}$^\dagger$ & \textbf{0.126}$^\dagger$
& \textbf{0.941}$^\dagger$ & \textbf{0.711}$^\dagger$
& \textbf{0.849}$^\dagger$ & \textbf{0.465}$^\dagger$
& \textbf{0.826}$^\dagger$ & \textbf{0.490}$^\dagger$ \\

\midrule

\multirow{2}{*}{Hi-BEHRT}
& $\times$
& 0.620 & 0.069
& 0.886 & 0.542
& 0.798 & 0.392
& 0.703 & 0.298 \\
& $\surd$
& \textbf{0.684}$^\dagger$ & \textbf{0.100}$^\dagger$
& \textbf{0.934}$^\dagger$ & \textbf{0.691}$^\dagger$
& \textbf{0.835}$^\dagger$ & \textbf{0.443}$^\dagger$
& \textbf{0.786}$^\dagger$ & \textbf{0.401}$^\dagger$ \\

\bottomrule
\end{tabular}
}
\label{tab:w/wo}
\vspace{-2mm}
\end{table*}

\section{Main Results}

\paragraph{MIMIC-IV Results.}
Table~\ref{tab:mimic_results} reports the performance of \texttt{EHR-RAGp} and the evaluated baselines on the four MIMIC-IV dataset tasks. \texttt{EHR-RAGp} achieves the highest AUROC and AUPRC across all tasks. Relative to the strongest baseline for each metric, it improves AUROC by 0.029 for ICU readmission, 0.021 for in-hospital mortality, 0.021 for long LOS, and 0.047 for 1-year mortality. The corresponding AUPRC improvements are 0.013, 0.060, 0.044, and 0.091, respectively. Paired-bootstrap tests indicate that the differences from the strongest baseline are statistically significant for all metrics except ICU-readmission AUPRC, for which the improvement over GenHPF is not significant. Beyond the strongest baselines, \texttt{EHR-RAGp} consistently achieves higher performance gain across both medical and transformer based baselines. Table~\ref{tab:llm_baselines} shows that LLM baselines perform substantially worse across all tasks. Overall, these results suggest that retrieving relevant information from the longitudinal patient history can complement fixed-context representations across various clinical outcomes. 


\vspace{-2mm}
\paragraph{EHRSHOT Results.}
Table~\ref{tab:ehrshot_results} compares \texttt{EHR-RAGp} with CLMBR and the long-context models on the EHRSHOT tasks. Relative to its CLMBR backbone, \texttt{EHR-RAGp} yields positive AUROC differences across all nine tasks. Among the reported results, \texttt{EHR-RAGp} obtains the highest point estimate on eight of all tasks. Overall, \texttt{EHR-RAGp} provides substantially larger gains on the new-diagnosis assignment tasks than on the operational outcomes. Moreover, \texttt{EHR-RAGp} matches or exceeds the reported long-context models on eight of the nine EHRSHOT tasks and remains competitive on the remaining task. These findings suggest that retrieval can preserve, and frequently improve upon, the performance of long-context modeling without relying on an increasingly large context window.




\vspace{-3mm}
\paragraph{\texttt{EHR-RAGp} is Model-Agnostic.} Table~\ref{tab:w/wo} demonstrates the benefit of integrating \texttt{EHR-RAGp} with other model backbones. Across the four models, tasks, and evaluation metrics, incorporating \texttt{EHR-RAGp} improves performance, yielding positive differences in all 32 model/task/metric comparisons. The AUROC gains range from 0.008 to 0.094, while the AUPRC gains range from 0.017 to 0.149. Particularly large improvements are observed for Med-BERT on 1-year mortality, where AUROC increases from 0.729 to 0.823 and AUPRC from 0.338 to 0.476, and for Hi-BEHRT on in-hospital mortality, where AUPRC increases from 0.542 to 0.691. Paired-bootstrap testing indicates that 31 of the 32 improvements are statistically significant; the only exception is the AUPRC improvement of CEHR-BERT on long LOS. These results, along with CLMBR results on EHRSHOT, demonstrate that the gains provided by \texttt{EHR-RAGp} are not restricted to a single backbone and that retrieving relevant longitudinal history consistently strengthens a range of existing EHR models.

\begin{wraptable}{r}{0.5\textwidth}
\vspace{-4mm}
\caption{{Ablation of retrieval mechanisms. $^\dagger$denotes p$<$ 0.05 compared to \texttt{EHR-RAGp}.} }
\resizebox{.9\linewidth}{!}{
\begin{tabular}{ccccccccc}
\toprule
\multirow{2}{*}{\textbf{Config.}} 
& \multicolumn{2}{c}{\textbf{Long LOS 7d}} 
& \multicolumn{2}{c}{\textbf{1YR Mortality}} \\

& \textbf{AUROC} & \textbf{AUPRC} 
& \textbf{AUROC} & \textbf{AUPRC} \\
\midrule

w/o Retrieval   
& 0.841$^\dagger$ & 0.465$^\dagger$
& 0.805$^\dagger$ & 0.466$^\dagger$ \\ 

w/o Prototypes     
& 0.859$^\dagger$  &  0.491$^\dagger$
& 0.839$^\dagger$  & 0.516$^\dagger$\\
 \midrule
\texttt{EHR-RAG$_p$} 
& \textbf{0.867} &  \textbf{0.509} 
& \textbf{0.853} & \textbf{0.564} \\
\bottomrule

\end{tabular}}
\label{tab:retrieval_mech}
\vspace{-4mm}
\end{wraptable}

\vspace{-1mm}
\section{Ablations}
\vspace{-2mm}
\paragraph{Effect of Prototypes.} 

Table~\ref{tab:retrieval_mech} separates the contributions of retrieving longitudinal history and prototype-guided alignment. Relative to the RoBERTa model without retrieval, incorporating semantically retrieved history without prototypes improves AUROC and AUPRC by 0.018 and 0.026, respectively, for long-LOS prediction, and by 0.034 and 0.050 for 1-year mortality. Introducing prototype-guided alignment provides further improvements of 0.008 AUROC and 0.018 AUPRC for long LOS, and 0.014 AUROC and 0.048 AUPRC for 1-year mortality. Differences between the full model and the ablated variants are statistically significant under paired-bootstrap testing. These results suggest that access to longitudinal history provides a consistent benefit, while prototypes alignment contributes additional improvements, particularly in AUPRC.

\newpage
\begin{wraptable}{r}{0.48\textwidth}
\centering
\vspace{-4.8mm}
\caption{Ablation of the usage of regularization.}
\resizebox{0.9\linewidth}{!}{
\begin{tabular}{ccccccc}
\toprule
\multirow{2}{*}{\bm{$\lambda_u$}}
& \multicolumn{2}{c}{\textbf{Query}}
& \multicolumn{2}{c}{\textbf{History}}
& \multicolumn{2}{c}{\textbf{Performance}} \\
& \textbf{p Max}
& \textbf{Entropy}
& \textbf{p Max}
& \textbf{Entropy}
& \textbf{AUROC}
& \textbf{AUPRC} \\
\midrule
\multicolumn{7}{c}{\textbf{Long LOS 7d}} \\
\midrule
$\times$
& 0.887 & 0.663
& 0.094 & 4.048
& 0.843 & 0.472\\
$\checkmark$
& 0.084 & 4.096
& 0.027 & 4.156
& \textbf{0.867} & \textbf{0.509} \\
\midrule
\multicolumn{7}{c}{\textbf{ICU Readmission 30d}} \\
\midrule
$\times$
& 0.716 & 1.578
& 0.030 & 6.099
& 0.715 & 0.118 \\
$\checkmark$
& 0.014 & 6.198
& 0.004 & 6.232
& \textbf{0.750} & \textbf{0.141} \\
\bottomrule
\end{tabular}
}
\vspace{-1mm}
\label{tab:Reg-effect}
\end{wraptable}

\textbf{Effect of Regularization.} Table~\ref{tab:Reg-effect} evaluates the impact of prototype regularization ($\lambda_u$) on both representation behavior and downstream performance. Without regularization, prototype assignments are highly peaked for queries (e.g., $P_{\max}=0.887$ for Long LOS), indicating collapse to a few dominant prototypes, while history assignments remain sparse and less informative. Enabling regularization significantly increases entropy and reduces $P_{\max}$ for both query and history, leading to more distributed and balanced prototype usage. This results in consistent performance gains across tasks, with Long LOS AUPRC improving from $0.472$ to $0.509$, and ICU readmission AUPRC from $0.118$ to $0.141$. Overall, the results show that regularization prevents prototype collapse, promotes better utilization of the prototype space, and improves performance. See Appendix \ref{sec:prot-reg-qual} for further analysis on the usage regularization $\lambda_u$.

\begin{wraptable}{r}{0.50\textwidth}
\vspace{-4.5mm}
\centering
\caption{Impact of the chunking strategy.}
\label{tab:chunking_acute}
\resizebox{0.9\linewidth}{!}{
\begin{tabular}{lcccc}
\toprule
\textbf{Model}
& \multicolumn{2}{c}{\textbf{ICU Readmission 30d}}
& \multicolumn{2}{c}{\textbf{In-Hospital Mortality}} \\
& \textbf{AUROC} & \textbf{AUPRC}
& \textbf{AUROC} & \textbf{AUPRC} \\
\midrule

Event-Based
& \textbf{0.750} & \textbf{0.141}
& \textbf{0.950} & 0.736 \\

Time-Based
& 0.741 & 0.129
& 0.946 & 0.738 \\

Visit-Level
& 0.748 & 0.140
& 0.948 & 0.738 \\

Care-Stage
& 0.747 & 0.138
& 0.947 & \textbf{0.739} \\

\bottomrule
\end{tabular}
}
\vspace{-.8mm}
\label{tab:chunking1}
\end{wraptable}

\textbf{Effect of Chunking.} 
Table~\ref{tab:chunking1} compares the four chunking strategies on ICU readmission and in-hospital mortality. Performance varies only slightly across the strategies. Event-based chunking achieves the highest ICU-readmission AUROC and AUPRC ($0.750$ and $0.141$) and the highest in-hospital mortality AUROC ($0.950$), while care-stage chunking obtains the highest in-hospital mortality AUPRC ($0.739$). Time- and visit-based chunking remain competitive across both tasks. Overall, these results indicate limited sensitivity to the choice of chunking strategy, with modest task-dependent differences. 

Additional results and ablations are provided in  Appendix~\ref{sec:additional_results}.

\vspace{-2mm}

\section{Discussion \& Conclusion}
\vspace{-3mm}
In this work, we introduced \texttt{EHR-RAGp}, a retrieval-augmented framework that incorporates extended longitudinal patient history through semantic retrieval and prototype-guided alignment. Across two benchmarks and thirteen prediction tasks, \texttt{EHR-RAGp} achieves the strongest results across all MIMIC-IV metrics, improves upon its CLMBR backbone on all EHRSHOT tasks, and matches or exceeds the reported long-context models on eight of nine tasks. Integrating the framework with four additional EHR encoders also improves all 32 model/task/metric combinations, with 31 reaching statistical significance. Controlled ablations indicate that both retrieved history and prototype-guided alignment contribute to performance, while prototype regularization promotes more balanced prototype usage. Performance remains relatively consistent across the four chunking strategies, suggesting limited sensitivity to how patient histories are segmented. Overall, the results support retrieval as a useful complement to fixed-context EHR representations, particularly for outcomes that depend on information distributed across the longitudinal record.

\paragraph{Limitations.} First, although we evaluate on two EHR benchmarks, additional institutions, patient populations, and coding systems are needed to assess broader generalizability. Second, \texttt{EHR-RAGp} focuses on structured EHR events, while patient records include other modalities such as clinical notes, images, and physiological waveforms. Future work could extend retrieval across these data sources. Finally, candidate retrieval relies on a fixed pretrained encoder and precomputed similarity index. Jointly learned or clinically informed retrieval represents a promising direction for improving task-specific adaptation.

\newpage
\bibliography{mybibliography}
\bibliographystyle{unsrt}
\newpage
\appendix
\section{Appendix}

\setcounter{figure}{0}
\setcounter{table}{0}

\renewcommand{\thefigure}{S\arabic{figure}}
\renewcommand{\thetable}{S\arabic{table}}

\section{Data Preprocessing}

We describe the preprocessing details performed on MIMIC-IV dataset in this section from raw tables to training ready data. For EHRSHOT dataset, the maintainer provides a training ready dataset along with splits and a labels.  For further details regarding EHRSHOT, we refer the reader to the original work \citep{wornow2023ehrshot}. 
\label{sec:preprocessing}

\subsection{Cohort construction}
\label{sec:cohort}
\paragraph{Pretraining Cohort:}
We first convert the MIMIC-IV dataset into the Medical Event Data Standard (MEDS) format \citep{arnrich2024medical, mcdermott2025meds}, representing each patient history as a chronologically ordered sequence of medical events covering all patient visits within a single sequence. MEDS extraction pipelines of MIMIC-IV fix the splits percentages upon extraction to $80\%$ train, $10\%$ tuning, and $10\%$ held-out based on the \textit{subject\_id}. We keep the splits percentages as is across the whole preprocessing pipelines for ease of data reproducibility. Following that, we use the training split as the pretraining cohort in our work and exclude both tuning and held-out splits to avoid any possible \textit{leakage}. The resulting cohort consists of  $291,702$ patients with $707,716,112$ medical events. Finally, we hold out $5\%$ of the training patients for pretraining quality monitoring.

\paragraph{Downstream Cohort:}
For downstream tasks, we construct the dataset using ICU stays only and exclude patients with hospital admissions but no ICU stays. We follow the inclusion and exclusion criteria proposed in previous work \citep{harutyunyan2019multitask} to define the downstream cohort. First, we include all adult patients with an age at admission of $\geq 18$ years. Second, we retain patients with a single ICU stay within the same hospital admission. Third, we include only patients whose admission and discharge ICU units are identical. Finally, we examine the length of each valid ICU stay and retain only those stays lasting at least $24$ hours. This process results in a downstream cohort comprising $49{,}839$ patients with $61{,}163$ ICU stays.

\subsection{Data Preprocessing}
We perform preprocessing to reduce irregularities and transform the data into a standardized format. First, we exclude HCPCS billing codes from the patient timelines, as they appear in the charted data in the patient timeline rather than at discharge time. This situation may introduce temporal leakage of billing codes, so we safely remove them. Second, we observed that vital signs are present in the \textit{OMR.csv} table as text values rather than numerical values. We convert them into numerical values and store them in the \textit{numeric\_value} column instead of the \textit{text\_value} column. Third, we handle outliers in numeric values using the \textit{MEDS-transform} Python package by computing dataset-wide statistics over the \textit{train} split for each numerical event and excluding values that lie beyond three standard deviations $(3\sigma)$ from the mean. Fourth, we filter out codes that appear rarely in the dataset, with an appearance frequency of $<4$ events. We also exclude events with no valid timestamp, except for static events that are intentionally without timestamps. Finally, we normalize all remaining numeric values using \textit{MEDS-transform} by recomputing normalization statistics after outlier removal.

\subsection{Patient Timeline Construction}
\label{sec:sequence-build}
\begin{figure}[t]
    \centering
    \includegraphics[width=\linewidth]{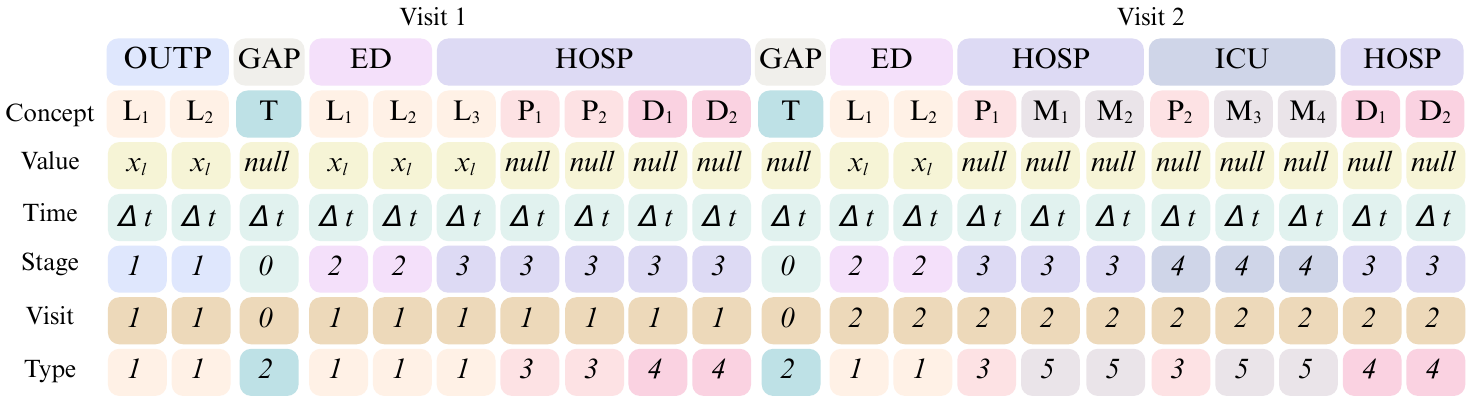}
    \caption{Sample patient timeline consisting of two visits. We illustrate the main components constituting the patient timeline in our implementation including medical events, numeric values, time values, care stage representation, visit order representation, and event type representation. Abbreviations, OUTP: Outpatient, ED: Emergency Department, HOSP: Hospital Admission, ICU: Incentive Care Unit, L: Labs, T: TIME-GAP, P: Procedure, M Medication, D: Diagnosis, $x_a$ numeric value. This is a sample timeline created randomly for illustration purposes and does not reflect real data.}
    \vspace{-2mm}
    \label{fig:sample-seq}
\end{figure}

Figure~\ref{fig:sample-seq} depicts a sample patient timeline segmented into distinct care stages and covering all relevant clinical information. To construct a patient timeline, we first examine each patient sequence and identify hospital admissions (HOSP) using the \textit{hadm\_id} field, as well as ICU stays  using the \textit{icustay\_id} field, which are the only care stages in MIMIC-IV associated with unique identifiers.

During this process, we observe that certain medical events, appear in the patient timeline without an explicit reference to any hospital admission. To handle these events without excluding them, we assess their temporal proximity to admissions present in the patient timeline. Events occurring after \textit{ED\_registration}, and events up to \textit{HOSPITAL\_ADMISSION}, are labeled as emergency department (ED) events, while events that occur between two consecutive hospital stays outside the admission/emergency boundaries are treated as outpatient (OUTP) events.

To represent time gaps between consecutive visits, we insert artificial time tokens following prior work \citep{pang2021cehr,fallahpour2024ehrmamba}, denoted as GAP, as illustrated in Figure~\ref{fig:sample-seq}. These tokens encode temporal intervals spanning, days, weeks, months, or years, depending on the elapsed time between visits.

A standard patient timeline primarily consists of hospital events including laboratory measurements, medications, ICD procedure codes, ICD diagnosis codes, and DRG codes, as well as ICU events such as chart events, infusions, procedures, and fluid outputs.

Each patient sequence is prepended with two events indicating the gender and birth date. While gender is a static event with no timestamp, birth date is a timestamped event represented as \textit{MEDS\_BIRTH}. If a patient dies during or after discharge, a special timestamped token, \texttt{MEDS\_DEATH}, is appended to the sequence. We also include special tokens required for language modeling objectives, such as \texttt{[PAD]}, \texttt{[MASK]}, \texttt{[CLS]}, and \texttt{[UNK]}, where applicable. 

For events associated with numeric values, such as laboratory tests and ICU chart events, we attach the corresponding numeric measurements, normalized. For events without numeric values, including \texttt{PROCEDURE-ICD} and \texttt{DIAGNOSIS-ICD}, we associate a learnable \textit{null} parameter. To encode temporal information, we compute the time difference between consecutive events in minutes and attach this value to each event. Because time gaps may vary substantially both within and across visits, we scale time deltas using the transformation defined in Equation~\ref{eqn:time-scale}, which maps values to the range $[0,1]$ and prevents extreme magnitudes. 

\begin{equation}
    \label{eqn:time-scale}
    \Delta t_i^{'} = \frac{log(1 + \Delta t_i )}{log(\Delta T_{max})}
\end{equation}

where $\Delta t_i$ is the consecutive time difference at event $i$, and $\Delta T_{max}$ is the maximum time difference present in the dataset. Temporal representations are then processed using Time2Vec \citep{kazemi2019time2vec} during training.

To further contextualize each event within the patient timeline, we introduce three additional categorical representations, care stage, visit order, and event type, as illustrated in the bottom rows of Figure~\ref{fig:sample-seq}. The care stage embedding explicitly encodes the clinical context in which each event occurs, distinguishing between outpatient (OUTP), emergency department (ED), inpatient hospitalization (HOSP), ICU stay (ICU), and artificial gap (GAP) events. This representation enables the model to differentiate identical medical events occurring under different clinical settings, which often carry distinct semantic meanings.

In addition, we assign a visit order index to each event, indicating the chronological visit number to which it belongs. The visit order is incremented across care episodes and reset for time gap tokens, allowing the model to reason over longitudinal disease progression across multiple encounters rather than treating the timeline as a flat sequence. Finally, the event type embedding encodes the high-level category of each event, such as laboratory tests, medications, diagnoses, procedures, administrative markers, or special tokens. Below, we provide a formal definition of each component that define a complete event $(c_{nk}, v_{nk}, t_{nk})$:

\begin{enumerate}
    \item \textbf{Concept Embedding:} Each clinical concept $c_{nk}$ (e.g., diagnosis, lab test, procedure, medication) is associated with a learnable semantic embedding: $\mathbf{u}_{c_{nk}} \in \mathbb{R}^{d}$.
    
    \item \textbf{Value Embedding:} For events with a numerical measurement $v_{nk}$, the value is normalized and projected, such that $\mathbf{v}_{nk} = \theta(v_{nk}) \in \mathbb{R}^{d}$, and $\theta(\cdot)$ is a multi-layer perceptron. Events without values are assigned a learnable (null-value) parameter.

    \item \textbf{Time Encoding:} Each timestamp $t_{nk}$ is mapped into a temporal embedding representing the local temporal information as time deltas between consecutive events, such that $\mathbf{t}_{nk} \in \mathbb{R}^{d}$. 

    \item \textbf{Visit Order Embedding:} To capture patient-level longitudinal ordering, each event inherits a visit index embedding: $\mathbf{r}_{\text{visit}(n)} \in \mathbb{R}^{d}$, where the indices are encoded as continuous learnable positional vectors. This allows the model to distinguish between early and late episodes in the patient trajectory (global temporal representation).

    \item \textbf{Care Stage Embedding:} Each visit belongs to a clinical stage (e.g., outpatient, emergency, inpatient, ICU). We assign a learned embedding: $\mathbf{s}_{\text{stage}(n)} \in \mathbb{R}^{d}$ to enable the model to incorporate care-context information, e.g.,  ICU vs. emergency events carry different clinical semantics.

    \item \textbf{Type Embedding:} Each event also receives a higher-level event type identifier that encodes the data source from which it originates (ICU chart event, ICU fluid output, etc.): $\mathbf{w}_{\text{type}(nk)} \in \mathbb{R}^{d}$. This provides local structural information that allows the model to differentiate heterogeneous event types.
\end{enumerate}

 To process these events within a unified architecture, we map each event and its associated metadata into a single representation $e_{nk} \in \mathbb{R}^d$, where $d$ is the vector dimension. This embedding considers six key components: clinical concept $\mathbf{u}_{c_{nk}}$, numeric value $\mathbf{v}_{nk}$, time $\mathbf{t}_{nk}$, visit-order $\mathbf{r}_{\text{visit}(n)}$,  care stage $ \mathbf{s}_{\text{stage}(n)}$, and event-type $\mathbf{w}_{\text{type}(nk)}$ embeddings as shown below. The unified event representation is then constructed by summation following \citep{fallahpour2024ehrmamba, li2020behrt}:
\begin{equation*}
    e_{nk} =(
            \mathbf{u}_{c_{nk}}
           +
            \mathbf{v}_{nk}
           +
            \mathbf{t}_{nk}
           +
            \mathbf{r}_{\text{visit}(n)}
           +
            \mathbf{s}_{\text{stage}(n)}
           +
            \mathbf{w}_{\text{type}(nk)})
         \in \mathbb{R}^d
\end{equation*}

By jointly modeling care stage, visit order, and event type, the patient timeline representation preserves structural, temporal, and semantic distinctions across events, facilitating more expressive and context-aware representation learning over heterogeneous EHR sequences.

\subsection{Query/History Definition}
\label{sec:query/hist}
\begin{figure}[!ht]
    \centering
    \includegraphics[width=\linewidth]{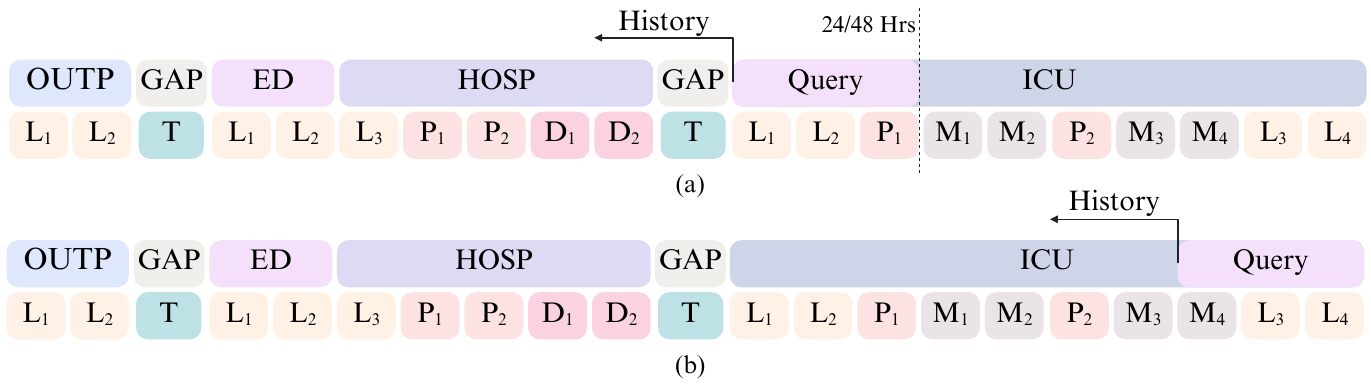}
    \caption{Illustration of query and history definition (a): Query/history boundaries for tasks with time-bounded prediction window. (b) Query/history boundaries task with entire stay prediction window.}
    \label{fig:query-history}
\end{figure}

For each prediction instance in the downstream cohort, we decompose the patient timeline into a \textit{query} segment and a corresponding \textit{history} segment to enable retrieval-based conditioning. The query represents the patient’s current clinical state exactly just before the time of prediction and consists of events observed within a task-specific observation window. For details on the observation windows used for each task, we refer the reader to Section~\ref{sec:tasks} and Table~\ref{tab:task_summary}. All events occurring strictly before the query window constitute the patient history and are segmented into fixed-length units that serve as retrievable candidates.  Figure~\ref{fig:query-history} illustrates the query/history definitions for the various prediction windows considered in this work. This formulation mirrors real-world clinical reasoning, in which decisions are made based on the present encounter while selectively consulting relevant past visits. By explicitly separating query and history, the model is encouraged to retrieve and integrate the most relevant historical context, rather than uniformly encoding the entire past, enabling scalable reasoning over long and heterogeneous patient trajectories.

\subsection{Leakage Prevention}

Due to the sensitivity of this query/history definition with respect to temporal leakage, we take strict measures to avoid any leakage. (1) We first index each patient timeline of $n$ events from $0-n$. (2) We then specify the query boundaries on the indexed patient timeline. (3) We examine each query if its contents could include any billing codes that might leak labels. (4) We iteratively refine query boundaries within the task window boundaries to ensure absence of any ineligible information. The same process is applied to the history chunks, especially those chunks closer to the query boundaries.

\subsection{Chunking}

\begin{figure}
    \centering
    \includegraphics[width=0.9\linewidth]{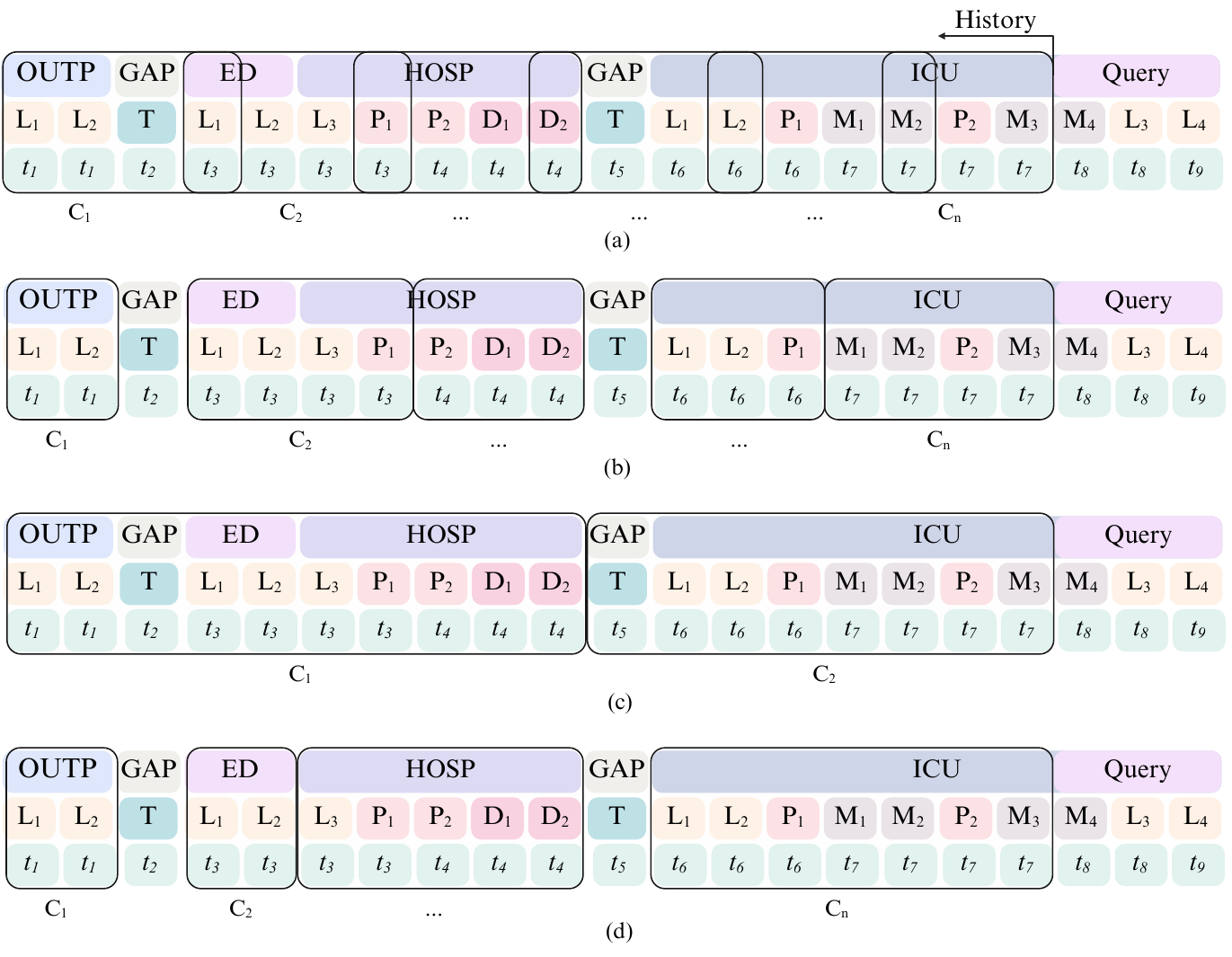}
    \caption{Visual illustration of the chunking strategies employed with \texttt{EHR-RAGp}. (a) Event-based chunking. (b) Time-based chunking. (c) Visit-level chunking. (d) Care-stage chunking. }
    \label{fig:chunking-methods}
\end{figure}

Structured EHR trajectories are inherently irregular, with clinical events occurring at variable temporal resolutions and across diverse care settings. To organize longitudinal patient history into retrievable units, \texttt{EHR-RAGp} partitions the timeline into clinically coherent chunks.  We investigate four chunking strategies: \textit{event-based}, \textit{time-based}, \textit{visit-level}, and \textit{care-stage} chunking. These strategies provide complementary views of patient history by capturing different temporal and clinical granularities. Figure \ref{fig:chunking-methods} visually illustrates these chunking strategies. Below, we formally outline our proposed chunking strategies:

\begin{itemize}

    \item \textbf{Event-Based:} Events are segmented into fixed-size groups based on event count. Each chunk contains a predefined number of embedded clinical events. Consecutive chunks overlap by a fixed number of events to preserve continuity across chunk boundaries. We set the chunk overlap to $12.5\%$ of the original chunk length.
    
    \item \textbf{Time-Based:} Events are segmented into fixed time windows, for example 6-hour, 12-hour, 24-hour blocks. Each chunk contains all events occurring in its temporal window.
    
    \item \textbf{Visit-level:} Each hospital visit, admission, or encounter is treated as a candidate chunk. This aligns with natural clinical boundaries: emergency department stays, inpatient admissions, and outpatient encounters.
    
    \item \textbf{Care-Stage:} Events are grouped by clinical stage (outpatient, emergency, inpatient, ICU), ensuring that each chunk reflects a coherent care stage.
\end{itemize}

In cases where the chunk length is shorter than the context window, we pad the sequence to match the context length. In cases where the chunk length is greater than the context length, we recursively impose \textit{Event-Based Chunking} with proper metadata handling, without overlap.

\vspace{-2mm}

\section{Experimental Setup}
\label{sec: exp-setup}

\subsection{Data Splitting}
\label{sec:splits}
We use a fixed train/validation/test partition shared across all downstream prediction tasks. Of the 61,163 eligible stays, 49,080 (80.2\%) are assigned to training, 5,986 (9.8\%) to validation, and 6,097 (10.0\%) to testing. The validation set is used for model selection and hyperparameter tuning, while the test set is reserved for final evaluation. Table~\ref{tab:splits} summarizes the task-specific label distributions within each split. To prevent information leakage, all stays associated with the same patient are assigned to a single split. The splits are inherited from the initial MEDS extraction, for ease of reproducibility.

\subsection{Downstream Tasks}
\label{sec:tasks}
\begin{itemize}
    \item \textbf{In-Hospital Mortality:} This task aims to predict whether a patient will die during the current hospital admission as a binary classification problem, using information available up to a predefined observation window. It is a widely used benchmark for evaluating clinical risk prediction models. In our setup, we condition the in-hospital mortality task on the first $48$ hours spent in the ICU, following \citep{harutyunyan2019multitask,hayat2022medfuse,elsharief2025medmod}.
    
   \item \textbf{1-Year Mortality Post-Discharge:} This task focuses on predicting whether a patient will die within one year following hospital discharge or no. It captures longer-term outcomes and reflects the model’s ability to leverage both acute and chronic patterns in a patient’s longitudinal history. We consider the entire ICU stay data as a valid prediction window \citep{meng2024machine,nistal2022developing,ghassemi2014unfolding}. For patients with multiple ICU stay and positive label of mortality within $1$-year, we only assign the positive label to all stays that are within one year from the mortality date.
    
    \item \textbf{Long Length of Stay:} This task predicts whether a patient’s hospital stay will exceed a predefined duration threshold. It serves as a proxy for resource utilization and clinical complexity and is commonly used to assess a model’s ability to identify patients at risk of prolonged hospitalization. We set the length-of-stay threshold to $7$ days following \citep{wornow2024context,wornow2023ehrshot,kim2023general} and define the prediction window as the first $24$ hours spent in the ICU, following \citep{zhang2024early,alghatani2021predicting}.

    \item \textbf{ICU Readmission:} This task aims to predict whether a patient will be readmitted to the intensive care unit within a specified time window after ICU discharge. It is an important indicator of care quality and patient instability and requires effective modeling of prior clinical trajectories. We set the readmission window to $30$ days post-discharge following \citep{hu2025prognostic,wornow2023ehrshot,chen2022machine} and use the entire ICU stay as the prediction window, following \citep{wornow2023ehrshot,de2023explainable}.
\end{itemize}

\begin{table*}[!ht]
    \caption{\small{Overview of the downstream tasks and the query/ history definition for each task}}
    \centering
    \resizebox{\linewidth}{!}{
    \begin{tabular}{l c c c c c} \toprule

        \textbf{Task}&  \textbf{Label type} & \textbf{Prediction Window}&  \textbf{Reference point} & \textbf{Query Window} &  \textbf{History Chunk} \\  

        \midrule
        \textbf{In-hospital mortality}  &  Binary &  $48$ hours since ICU admission & ICU admission & Last $L$ event $\leq$ ICU admission $+\ 48$ hours  & All events $<$  query start \\ 
        \textbf{Long Length of stay-7 Days}  & Binary  & $24$ hours since ICU admission & ICU admission & Last $L$ event $\leq$ ICU admission $+\ 24$ hours &  All events $<$  query start \\ 
        \textbf{ICU Readmission-30 Days}  & Binary & Entire ICU stay & ICU discharge & Last $L$ event $\leq$ ICU discharge & All events $<$  query start \\ 
        \textbf{1 Year Mortality}  & Binary & Entire ICU stay & ICU discharge & Last $L$ event $\leq$ ICU discharge  &  All events $<$  query start \\ 
\bottomrule
    \end{tabular}}
    \label{tab:task_summary}
\end{table*}

\begin{table}[!t]
\centering
\caption{Label distribution across data splits for the downstream prediction tasks.}
\label{tab:label-distribution-splits}
\resizebox{\linewidth}{!}{
\begin{tabular}{lcccccc}
\toprule
\textbf{Task}
& \multicolumn{2}{c}{
    \shortstack{\textbf{Train}\\($N=49{,}080$ stays)}
}
& \multicolumn{2}{c}{
    \shortstack{\textbf{Validation}\\($N=5{,}986$ stays)}
}
& \multicolumn{2}{c}{
    \shortstack{\textbf{Test}\\($N=6{,}097$ stays)}
} \\
\cmidrule(lr){2-3}
\cmidrule(lr){4-5}
\cmidrule(lr){6-7}
& \textbf{Pos. (\%)}
& \textbf{Neg. (\%)}
& \textbf{Pos. (\%)}
& \textbf{Neg. (\%)}
& \textbf{Pos. (\%)}
& \textbf{Neg. (\%)} \\
\midrule

\textbf{In-Hospital Mortality}
& 5,004 (10.196)
& 44,076 (89.804)
& 642 (10.725)
& 5,344 (89.275)
& 621 (10.185)
& 5,476 (89.815) \\

\textbf{Long LOS ($\geq 7$ Days)}
& 6,678 (13.606)
& 42,402 (86.394)
& 840 (14.033)
& 5,146 (85.967)
& 763 (12.514)
& 5,334 (87.486) \\

\textbf{1-Year Mortality Post Discharge}
& 8,271 (16.852)
& 40,809 (83.148)
& 1,029 (17.190)
& 4,957 (82.810)
& 1,029 (16.877)
& 5,068 (83.123) \\

\textbf{ICU Readmission within 30 Days}
& 2,093 (4.264)
& 46,987 (95.736)
& 227 (3.792)
& 5,759 (96.208)
& 273 (4.478)
& 5,824 (95.522) \\
\bottomrule
\end{tabular}
}
\label{tab:splits}
\end{table}

\subsection{\texttt{EHR-RAGp} Architecture}
\texttt{EHR-RAGp} uses a RoBERTa-base encoder \citep{liu2019roberta} with $12$ transformer layers, $12$ attention heads, a hidden size of $768$, and learned absolute positional embeddings. The encoder supports sequences of up to $512$ tokens and produces contextualized token representations, which are aggregated into sequence-level representations using either [CLS] or mean pooling.

Each event token is represented as the sum of multiple embedding components. In addition to the core event embedding, we incorporate categorical embeddings for care stage, visit order, and event type, all projected to the same hidden dimension ($768$) and learned jointly with the model. Numerical values associated with events are normalized and projected through a small MLP before being fused with the categorical embeddings. In the absence of a numeric value, we use a learnable null embedding of the same dimensionality. Temporal information is encoded using Time2Vec \citep{kazemi2019time2vec}, applied to scaled time-delta features and added to the token representation.

For retrieval, the pretrained backbone encoder is used to embed historical timeline chunks, which are stored in a vector database. During training, the query representation is used to retrieve the top-$M$ most similar history chunks based on cosine similarity in the embedding space. Retrieval is non-parametric and does not involve gradient updates during the lookup stage; however, the encoder \texttt{EHR-RAGp} itself is fine-tuned end-to-end during downstream training.

The prototype module consists of a learnable set of prototype vectors defined in the same embedding space as the query and history representations. Both query and retrieved history embeddings are projected into the prototype space, where relevance scores are computed via similarity between history embeddings and prototypes, conditioned on the query representation. These scores are used to weight or filter retrieved history chunks, enabling selective integration of long-range context.

The fusion module combines the query representation with the filtered history representations using a shallow transformer with $2$ layers and $4$ attention heads. The fused representation is then passed to a task-specific prediction head implemented as a lightweight feedforward network. For all downstream tasks, the prediction head outputs a single logit optimized using binary cross-entropy (BCE) loss. All architectural components beyond retrieval are differentiable and optimized jointly during supervised fine-tuning.

\subsection{Pretraining Details}
\label{sec:pretraining}
We adopt masked language modeling (MLM) objective applied to structured EHR event sequences for \texttt{EHR-RAGp} pretraining. During pretraining, patient timelines are tokenized into fixed-length sequences. We consider the full patient timeline during pretraining and pass it to the encoder backbone as overlapped chunks of $512$ tokens with $64$ tokens of overlap. Hence, we produce dense sequences, whereas padding occurs only in the last chunk to fulfill the sequence length. During pretraining, a subset of $15\%$ of the chunk events is randomly selected for masking following the RoBERTa \citep{liu2019roberta} masking strategy. Specifically, masked positions are replaced with a special [MASK] token with probability $80\%$, substituted with a random event token with probability $10\%$, or left unchanged with probability $10\%$. The model is trained to predict the original event identity at masked positions using a cross-entropy loss over the entire vocabulary of $48{,}506$ tokens.

 All embedding components, including event embeddings and categorical embeddings, are trained jointly during this stage. No temporal, or numeric value components are used during pretraining; our objective at this stage is solely to learn contextualized representations of structured EHR sequences. Pretraining is conducted on the full pretraining cohort described in Section~\ref{sec:cohort}.

We use \textit{AdamW} \citep{loshchilov2017decoupled} optimizer with an initial learning rate specified using cyclical learning rates following \citep{smith2017cyclical}, a weight decay of $1\times10^{-2}$, a batch size of $64$, and a cosine annealing learning scheduler. Training is performed for $100$ epochs over the pretraining corpus. We monitor pretraining quality using accuracy on a held-out validation set, as described in Section~\ref{sec:cohort}. Early stopping is implemented with a patience of three epochs if validation accuracy does not improve. The resulting pretrained backbone is subsequently reused for downstream fine-tuning and retrieval embedding generation. All pretraining experiments are conducted using a single NVIDIA A100 GPUs.

\subsection{Vector Database Setup}

To enable retrieval-augmented conditioning, we construct a patient-specific vector database that stores embeddings of historical visit segments derived from the pretrained \texttt{EHR-RAGp} backbone. We use Facebook AI Similarity Search  (FAISS\footnote{https://github.com/facebookresearch/faiss}) as the vector index for \texttt{EHR-RAGp}. For each patient in the downstream cohort, the history portion of the timeline (as defined in Section~\ref{sec:query/hist}) is segmented into fixed-length chunks, where each chunk corresponds according to the various chunking strategies proposed in our work. Each chunk is independently encoded using the pretrained backbone encoder, and the resulting pooled representation is stored as a dense vector.

We use mean pooling over the sequence representations and normalize the resulting embeddings prior to storage in the index. The vector index is constructed offline before downstream training and remains fixed during supervised fine-tuning. This design avoids repeated recomputation of historical embeddings and ensures scalable retrieval over long patient histories. During retrieval, we perform similarity search based on cosine similarity between the query embedding and historical chunk embeddings. It is worthy to note as well that we do not store chunked data along with their encoded vector, instead; we keep track of the chunk indices in the full chunked patient timeline. Upon retrieval, we perform online chunking during training to  retrieve selected candidates by similarity search via their indices. This approach allow for faster retrieval and reduce the need for maintaining large database that accommodates the chunked data along with the vector index.

\subsection{Downstream Training Details}
Table \ref{tab:hyperparams} summarizes the hyperparameter settings and search spaces used for downstream training of \texttt{EHR-RAGp}. Core backbone parameters are fixed across all experiments, including the RoBERTa-base encoder with a hidden dimension of $768$, $12$ layers, and $12$ attention heads.  

Retrieval-related hyperparameters including query chunk size, history chunk size, history chunk overlap, number of retrieved chunks, and similarity metric. all retrieval hyperparameters are held constant with values of $1024$, $256$, $32$, $24$, \textit{cosine similarity}, respectively. 

For prototypes module, we use fixed prototype dimensionality of $768$. For temperature, we experiment with temperature pairs, $(T_q, T_h)$, for prototype assignment of $[(0.025,0.1), (0.05,0.2),(0.02,0.08)]$ and  Weighing temperature $T_s$ values of $\{0.1,0.15,0.2,0.25\}$, whereas the number of prototypes is set to be optimized with discrete search space in $\{64\,128,256,512\}$. 

For fusion, we set up a transformer encoder of $2$ layers and $4$ attention heads, and dropout of $0.1$. We set the pooling strategy to be set during hyperparameter optimization with search space of $\{mean, query\}$. 

For training hyperparameters, we experiment with Stochastic Gradient Descent (SGD) optimizer, learning rate in $[1\mathrm{e}{-5}, 5\mathrm{e}{-4}]$, Weight decay in $[1\mathrm{e}{-3}, 1\mathrm{e}{-2}]$ and cosine annealing learning rate scheduler. For usage penalty $\lambda_u$, we experiment with fixed values in $\{0.004, 0.005, 0.006, 0.007\}$. 

For hyperparameters optimization, we run $25$  trails for each prediction task using Bayesian search. All downstream evaluation experiments were conducted using a single NVIDIA A100  80GB GPUs. Maximum batch size with cane be used with such a GPU in our experiments is $24$ query chunks of length $512$ events along with $24$ history chunk of length $256$, while average epoch time given the max settings is $38$ minutes. It is worthy to note that higher capacity GPUs can be also used with increase in the max settings mentioned earlier. With NVIDIA H200  141GB for instance, max batch size can reach $60$, while the epoch time can reach as low as $13$ minutes. Table \ref{tab:hyperparams} summarizes the the downstream tasks training hyperparameters for both fixed and optimized ones, whereas Table \ref{tab:best_hparams} reports the best hyperparameters obtained for each task.

\begin{table*}[t]
\centering
\small
\setlength{\tabcolsep}{6pt}
\caption{Hyperparameter settings and search spaces used in \texttt{EHR-RAGp} experiments.}
\resizebox{0.7\linewidth}{!}{
\begin{tabular}{llll}
\toprule
\textbf{Category} 
& \textbf{Hyperparameter} 
& \textbf{Fixed} 
& \textbf{Search Space} \\
\midrule

\multirow{5}{*}{Backbone}
& Encoder architecture & RoBERTa-base & -- \\
& Hidden dimension ($d$) & 768 & -- \\
& Number of layers & 12 & -- \\
& Number of attention heads & 12 & -- \\
& pooling & -- & \{mean, CLS\} \\
\midrule

\multirow{5}{*}{Retrieval}
& Query Chunk size  & 1024 & -- \\
& History Chunk size  & 256 & -- \\
& History Chunk overlap$^*$  & 32 & -- \\
& Number of retrieved chunks ($M$) & 24 & -- \\
& Similarity metric & Cosine & -- \\

\midrule

\multirow{5}{*}{Prototypes}
& Number of prototypes & -- & \{64, 128, 256, 512\} \\
& Prototype dimension & 768 & -- \\
& Prototype temperature $(T_q,T_h)$ & --& $\{(0.025,0.1), (0.05,0.2),(0.02,0.08)\}$  \\

& Weighing temperature $T_s$ & -- & $\{0.1,0.15,0.2,0.25\}$ \\

\midrule

\multirow{4}{*}{Fusion}
& Fusion layers & 2 & -- \\
& Fusion attention heads & 4 & -- \\
& pooling & -- & \{mean, query\} \\
& Dropout & 0.1 & -- \\

\midrule

\multirow{5}{*}{Training}
& Optimizer & SGD & -- \\
& Usage penalty $\lambda_u$ & -- & $\{0.004, 0.005, 0.006, 0.007\}$ \\
& Learning rate & -- & $[1\mathrm{e}{-5}, 5\mathrm{e}{-4}]$ \\
& Weight decay & -- & $[1\mathrm{e}{-3}, 1\mathrm{e}{-2}]$ \\
& Batch size & -- & \{8,12,16,20\} \\
& Max epochs & 75 & -- \\

\bottomrule
$^*$ overlap is used only with event-based chunking strategy \\
\end{tabular}
}
\label{tab:hyperparams}
\end{table*}

\begin{table*}[t]
\centering
\small
\caption{Best configurations selected for each downstream task after hyperparameter optimization.}
\begin{tabular}{lcccc}
\toprule
\textbf{Hyperparameter} &
\textbf{ICU-Readmit 30d} &
\textbf{In-hospital Mortality} &
\textbf{Long LOS 7d} &
\textbf{1YR Mortality} \\
\midrule
Chunking Strategy & Event-Based & Event-Based & Visit-Level & Care-Stage\\
Prototype count & $64$ & $256$ & $64$ & $512$ \\
$(T_q,T_h)$ & $(0.025,0.1)$ & $(0.05,0.2)$ & $(0.025,0.1)$ & $(0.05,0.2)$ \\
Weighting temperature ($T_s$) &  $0.25$ & $0.20$ & $0.25$ & $0.22$\\
Usage penalty ($\lambda_u$) & $0.005$ & $0.005$ & $0.006$ & $0.006$ \\
Learning rate & $2.592 \times 10^{-4}$ & $1.909 \times 10^{-4}$ & $1.119 \times 10^{-4}$ & $1.948 \times 10^{-5}$\\
Weight decay & $1.454 \times10^{-3}$ & $1.380 \times10^{-3}$ & $1.961 \times10^{-3}$ & $1.54 \times10^{-3}$\\
\bottomrule
\end{tabular}
\label{tab:best_hparams}
\end{table*}

\section{Baselines}
\label{sec:baselines}

\subsection{Clinical Baselines}

\begin{itemize}
    \item \textbf{DescEmb} \citep{hur2022unifying}: is a text-based framework that represents patient history as textual descriptions of medical events, covering the event type, numeric value (if any), along with its unit of measurement. These sequences are then processed via clinical or general-purpose pretrained language models. The main goal of DescEmb is to reduce reliance on predefined medical code vocabularies through a shared linguistic vocabulary that can be utilized across different EHR datasets. We consider two variants of DescEmb in our experiments, including \textit{BERT Fine-Tune} and \textit{CLS Fine-Tune}.
    
    \item \textbf{GenHPF} \citep{hur2023genhpf}: is an extension of DescEmb that goes beyond the basic description of the medical event type, value, and unit to include all features associated with an event in the raw EHR tables, enhancing medical event representation. For instance, a medication event will cover type, name, frequency, dosage, route, etc. It shares the same objective as DescEmb of being agnostic to differing EHR schemas across institutions.
    
    \item \textbf{Med-BERT} \citep{rasmy2021med}: is a BERT-based foundation model that is pretrained on structured EHR records for disease prediction tasks. It relies solely on structured diagnosis ICD codes as pretraining and evaluation data. The pretraining approach typically follows masked language modeling as proposed in the BERT paper \citep{devlin2019bert}.
    
    \item \textbf{BEHRT} \citep{li2020behrt}: is another BERT-based foundation model that learns contextual embeddings of the patient’s structured records in a similar fashion to Med-BERT via masked language modeling using diagnosis codes. However, it differs from Med-BERT by incorporating patient age as a temporal signal.
    
    \item \textbf{Hi-BEHRT} \citep{li2022hi}: is a hierarchical extension of BEHRT that aims to handle longer, more comprehensive patient records. It captures short- and long-term patient patterns from hierarchically segmented medical sequences and is pretrained with the self-supervised BYOL framework \citep{grill2020bootstrap}.
    
    \item \textbf{CEHR-BERT} \citep{pang2021cehr}: is another BERT-based model that incorporates temporal and visit information to improve disease prediction. It represents patient visits as sequences of medical codes, augmented with artificial time tokens, age embeddings, and time embeddings.
    
    \item \textbf{REMed} \citep{kim2023general}: is a retrieval-based foundation model that processes textual descriptions in a similar fashion to both DescEmb and GenHPF. It operates by randomly sampling events from patient history and ranking them to retrieve the most relevant ones for different downstream tasks. In contrast to our proposed model, REMed operates at the medical event level, whereas \texttt{EHR-RAGp} operates at historical segment level to cover broad history beyond limited set of events, 128 in case of REMed.
    \item \textbf{EHRMamba} \citep{fallahpour2024ehrmamba}: typically follows the embedding structure of CEHR-BERT but uses the state-space architecture MAMBA \citep{gu2024mamba} instead of a BERT encoder. The use of a state-space model aims to enable efficient processing of relatively long EHR sequences compared to transformer models.
\end{itemize}

\subsection{Transformer-based Baselines}
\begin{itemize}
    \item \textbf{RoBERTa} \citep{liu2019roberta}: is a transformer-based language model that builds upon BERT by removing the next-sentence prediction objective and adopting dynamic masking and larger pretraining corpora. RoBERTa proves that careful optimization of the pretraining pipleines significantly improves contextual representation learning, which makes it a strong general-purpose encoder baseline.
    
    \item \textbf{LongFormer} \citep{beltagy2020longformer}: is a transformer variant designed to handle long input sequences efficiently through a sparse attention mechanism that combines local windowed attention with task-specific global attention. This design reduces the quadratic complexity of standard self-attention, which in turn enables modeling of substantially longer contexts compared to standard transformers.
    
    \item \textbf{BigBird} \citep{zaheer2020big}: is another long-context transformer architecture that replaces full self-attention with a sparse attention pattern composed of random, local, and global attention. BigBird is theoretically proven to be a universal approximator while enabling scalable processing of long sequences, which making it suitable for tasks that involve extended contextual dependencies.
    
    \item \textbf{RoFormer} \citep{su2024roformer}: is a transformer model that introduces rotary positional embeddings (RoPE) to encode relative positional information directly into the self-attention mechanism. This approach improves extrapolation to longer sequences and better captures relative token relationships, which makes it effective for sequence modeling tasks with variable-length inputs.
    
    \item  \textbf{ModernBERT} \citep{warner2025smarter}: is a re-engineered BERT-style encoder that incorporates recent architectural and training improvements, including optimized attention implementations, improved normalization, and enhanced efficiency on modern hardware. ModernBERT aims to retain the strengths of encoder-based transformers while achieving better performance and scalability.
\end{itemize}

\subsection{Large Language Model Baselines}

\begin{itemize}
    \item \textbf{Qwen2.5-7B-Instruct } \citep{qwen2}: is a general-purpose instruction-tuned large language model designed for conversational reasoning and text understanding across diverse domains.
    \item \textbf{Mistral-7B-Instruct} \citep{jiang2023mistral}: is an instruction-following variant of the Mistral family of open-source large language models optimized for efficient inference and general reasoning capabilities. 
    \item \textbf{Llama-3.1-8B-Instruct}: \citep{grattafiori2024llama} is a general-purpose instruction-tuned large language model designed for conversational interaction, reasoning, and text understanding across diverse domains.
    \item \textbf{MedGemma-1.5-4B-it} \citep{sellergren2026medgemma} is a medically adapted instruction-tuned language model developed for healthcare and biomedical applications. It is designed to support clinical reasoning and medical text understanding, and is evaluated here as a medical LLM baseline.
    \item \textbf{BioMistral-7B} \citep{labrak2024biomistral}: is a biomedical domain-adapted variant of Mistral trained on large-scale biomedical corpora to improve performance on medical and scientific language tasks.
    \item \textbf{Llama3-OpenBioLLM-8B} \citep{OpenBioLLMs}: is a biomedical instruction-tuned large language model adapted from the Llama 3 architecture to improve medical knowledge, clinical reasoning, and biomedical text understanding.

\end{itemize}

\subsection{LLM Evaluation Protocol}
We evaluated both general-domain and medical-domain large language models using a zero-shot next-token prediction framework for clinical risk prediction. For each downstream task, patient timelines were converted into sequential textual event representations and truncated according to predefined prediction windows. To ensure fair comparison with encoder-based baselines, the same temporal slicing strategy and prediction horizons were used across all models. A fixed precomputed slices 1024 clinical events along with their numeric values events used. 

Each patient sequence was formatted as a chronological list of clinical events and passed directly to the language model without supervised fine-tuning. We used causal language modeling inference and extracted the logits corresponding to the ``Yes'' and ``No'' tokens from the next-token distribution. All other token logits were ignored, and probabilities were normalized over the binary ``Yes/No'' subset using a softmax operation. The normalized probability assigned to the ``Yes'' token was used as the prediction score. We present the prompt used for each task in  The prompts used for zero-shot evaluation are shown in Box~\ref{box:llm-prompts}.

Performance was evaluated using AUROC and AUPRC. Confidence intervals were estimated using nonparametric bootstrap resampling with 1000 iterations.

\begin{tcolorbox}[
    colback=gray!10,
    colframe=gray!40,
    title=LLM Prompt Templates,
    sharp corners,
    boxrule=0.4pt
]
\label{box:llm-prompts}
\scriptsize

\textbf{In-Hospital Mortality}

\begin{verbatim}
You are an expert clinical risk prediction model using electronic health records.

--- PATIENT DATA ---
Electronic Health Records:
{ehr_text}

--- TASK ---
Will this patient die during this hospital admission?

Answer only using one word: Yes or No.

Answer:
\end{verbatim}

\noindent\rule{\linewidth}{0.3pt}

\textbf{1-Year Mortality}

\begin{verbatim}
You are an expert clinical risk prediction model using electronic health records.

--- PATIENT DATA ---
Electronic Health Records:
{ehr_text}

--- TASK ---
Will this patient die within 1 year after hospital discharge?

Answer only using one word: Yes or No.

Answer:
\end{verbatim}

\noindent\rule{\linewidth}{0.3pt}

\textbf{ICU Readmission within 30 Days}

\begin{verbatim}
You are an expert clinical risk prediction model using electronic health records.

--- PATIENT DATA ---
Electronic Health Records:
{ehr_text}

--- TASK ---
Will this patient be readmitted to the ICU within 30 days after ICU discharge?

Answer only using one word: Yes or No.

Answer:
\end{verbatim}

\noindent\rule{\linewidth}{0.3pt}

\textbf{Long Length of Stay > 7 Days}

\begin{verbatim}
You are an expert clinical risk prediction model using electronic health records.

--- PATIENT DATA ---
Electronic Health Records:
{ehr_text}

--- TASK ---
Will this patient have a hospital length of stay longer than 7 days?

Answer only using one word: Yes or No.

Answer:
\end{verbatim}

\end{tcolorbox}

\section{Additional Results}
\label{sec:additional_results}
\subsection{LLM Baseline Results}
\label{sec:llm_baselines}

Table~\ref{tab:llm_baselines} reports the zero-shot performance of general-purpose and clinically adapted LLMs. Performance varies across tasks: MedGemma performs best on ICU readmission, BioMistral on long LOS, and Qwen2.5 on both mortality outcomes. Clinically adapted LLMs do not consistently outperform their general-purpose counterparts. As expected under a zero-shot setting, these models perform below the task-specific \texttt{EHR-RAGp} model; therefore, we include them as reference results rather than as directly matched baselines.

\begin{table*}[t]
\centering
\small
\setlength{\tabcolsep}{6pt}
\caption{{Performance results zero-shot evaluation across LLM baselines} }
\resizebox{\linewidth}{!}{
\begin{tabular}{lccccccccc}
\toprule
\multirow{1}{*}{\textbf{Model}} &
& \multicolumn{2}{c}{\textbf{ICU-Readmit 30d}} 
& \multicolumn{2}{c}{\textbf{In-hospital Mortality}} 
& \multicolumn{2}{c}{\textbf{Long LOS 7d}} 
& \multicolumn{2}{c}{\textbf{1YR Mortality}} \\

& & \textbf{AUROC \scriptsize(CI)} & \textbf{AUPRC \scriptsize(CI)} 
& \textbf{AUROC \scriptsize(CI)} & \textbf{AUPRC \scriptsize(CI)} 
& \textbf{AUROC \scriptsize(CI)} & \textbf{AUPRC \scriptsize(CI)} 
 & \textbf{AUROC \scriptsize(CI)} & \textbf{AUPRC \scriptsize(CI)} \\
 
\midrule
\multicolumn{10}{c}{\textbf{General-purpose LLMs}} \\
\midrule
Qwen2.5-7B-Instruct &
&  0.484 \scriptsize(0.451, 0.520) & 0.043 \scriptsize(0.038, 0.052) 
&  \textbf{0.729 \scriptsize(0.707, 0.750)} & \textbf{0.309 \scriptsize(0.273, 0.345)}
&  0.504 \scriptsize(0.483, 0.525) & 0.130 \scriptsize(0.119, 0.144)
&  \textbf{0.576 \scriptsize(0.557, 0.594)} & \textbf{0.201 \scriptsize(0.187, 0.219)} \\

Mistral-7B-Instruct &
& 0.470 \scriptsize(0.432, 0.504) & 0.044 \scriptsize(0.037, 0.054)
& 0.700 \scriptsize(0.678, 0.724) & 0.249 \scriptsize(0.219, 0.284) 
& 0.529 \scriptsize(0.507, 0.552) & 0.145 \scriptsize(0.131, 0.163) 
& 0.533 \scriptsize(0.513, 0.553) & 0.188 \scriptsize(0.173, 0.207) \\

Llama-3.1-8B-Instruct &
& 0.498 \scriptsize(0.463, 0.530) & 0.044 \scriptsize(0.038, 0.051) 
& 0.659 \scriptsize(0.636, 0.681) & 0.184 \scriptsize(0.162, 0.209)
& 0.452 \scriptsize(0.430, 0.474) & 0.113 \scriptsize(0.105, 0.124) 
& 0.520 \scriptsize(0.500, 0.537) & 0.177 \scriptsize(0.165, 0.192) \\

\midrule 
\multicolumn{10}{c}{\textbf{Clinical LLMs}} \\
\midrule
medgemma-1.5-4b-it &
& \textbf{0.515 \scriptsize(0.481, 0.546)} & \textbf{0.047 \scriptsize(0.040, 0.056)} 
& 0.531 \scriptsize(0.504, 0.555) & 0.117 \scriptsize(0.105, 0.132) 
& 0.531 \scriptsize(0.511, 0.552) & 0.142 \scriptsize(0.130, 0.159) 
& 0.496 \scriptsize(0.477, 0.516) & 0.169 \scriptsize(0.158, 0.183) \\

BioMistral-7B  &
& 0.466 \scriptsize(0.432, 0.498) & 0.041 \scriptsize(0.036, 0.048)
& 0.484 \scriptsize(0.461, 0.506) & 0.098 \scriptsize(0.089, 0.108)
& \textbf{0.554 \scriptsize(0.532, 0.576)} & \textbf{0.146 \scriptsize(0.134, 0.159)}
& 0.473 \scriptsize(0.453, 0.493) & 0.159 \scriptsize(0.148, 0.171) \\

Llama3-OpenBioLLM-8B &
& 0.498 \scriptsize(0.463, 0.528) & 0.043 \scriptsize(0.037, 0.050)
& 0.455 \scriptsize(0.430, 0.479) & 0.107 \scriptsize(0.094, 0.127) 
& 0.495 \scriptsize(0.472, 0.517) & 0.127 \scriptsize(0.116, 0.142) 
& 0.483 \scriptsize(0.465, 0.501) & 0.164 \scriptsize(0.151, 0.178) \\
\bottomrule
\vspace{-3mm}
\end{tabular}}
\label{tab:llm_baselines}
\end{table*}

\subsection{Latent Representation Structure}

\begin{figure}[!t]
\centering

\begin{subfigure}[t]{0.49\textwidth}
    \centering
    \includegraphics[width=\linewidth]{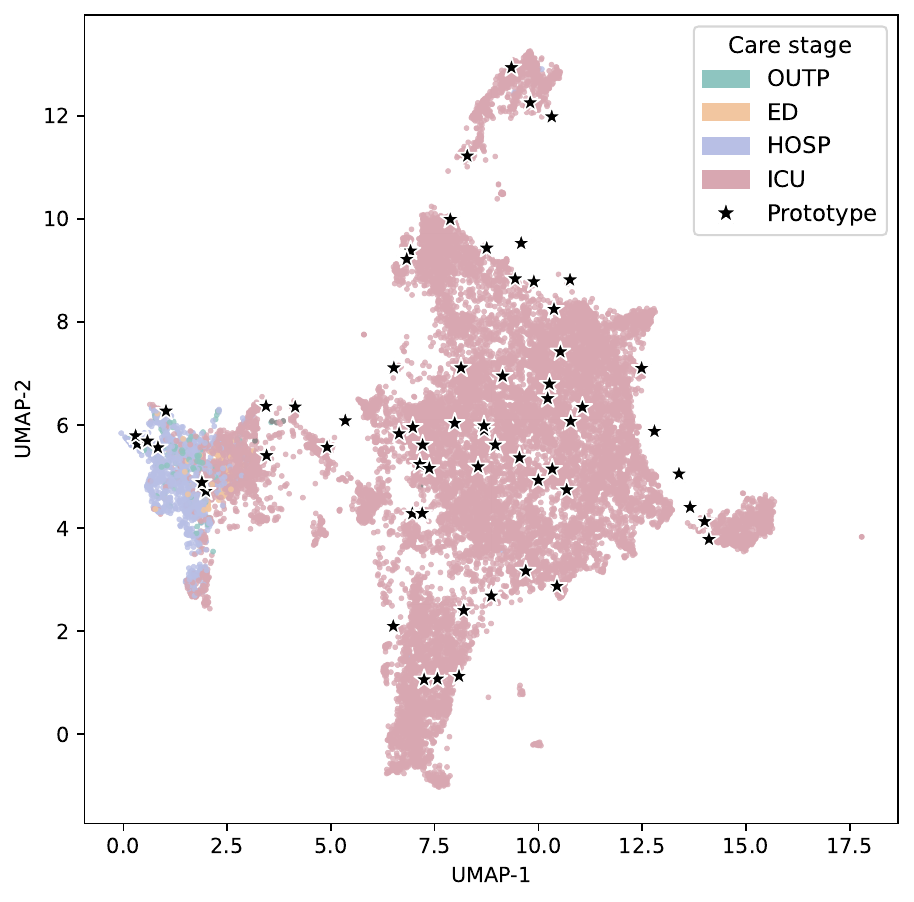}
    \caption{1-Year Mortality Post-Discharge}
    \label{fig:umap_mort_yr}
\end{subfigure}
\hfill
\begin{subfigure}[t]{0.49\textwidth}
    \centering
    \includegraphics[width=\linewidth]{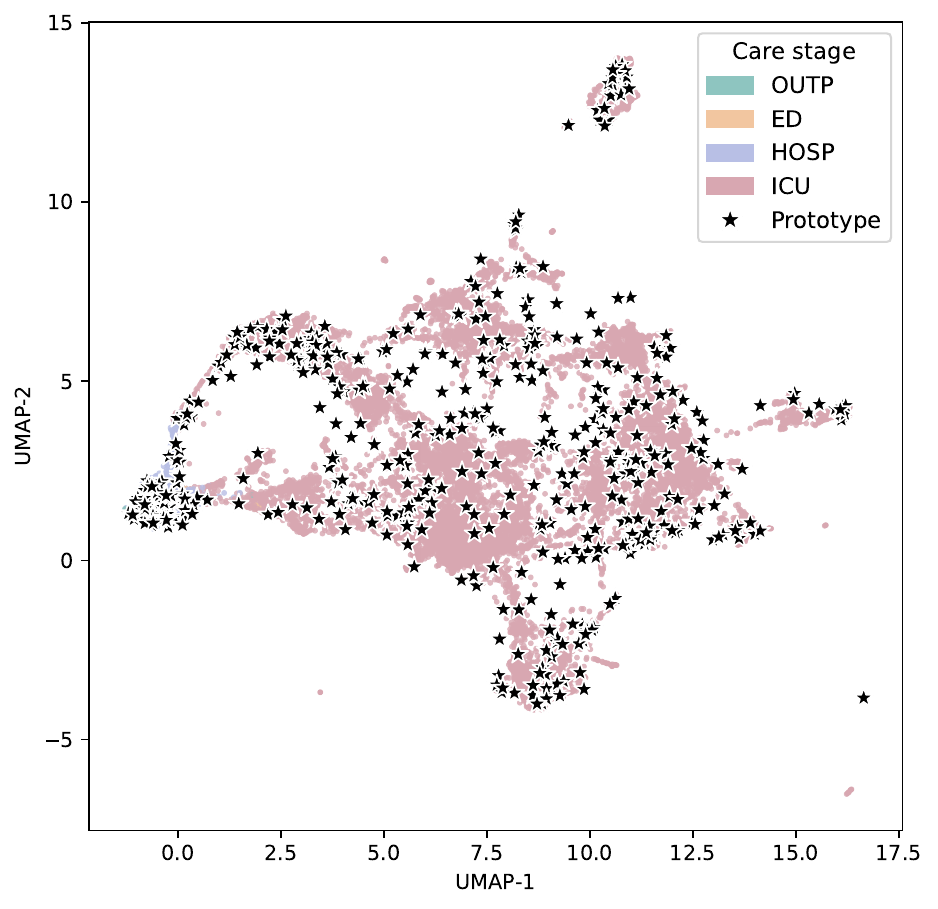}
    \caption{ICU Readmission within 30 Days}
    \label{fig:umap_icu}
\end{subfigure}

\par\vspace{4mm}

\begin{subfigure}[t]{0.49\textwidth}
    \centering
    \includegraphics[width=\linewidth]{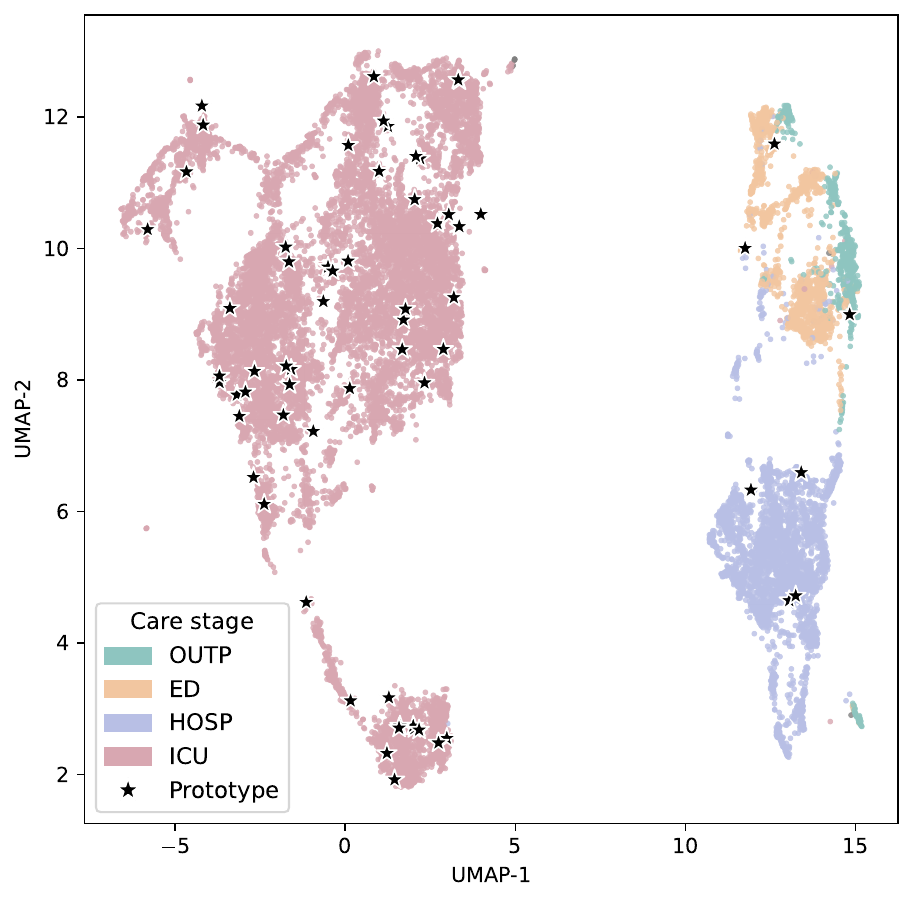}
    \caption{In-Hospital Mortality}
    \label{fig:umap_ih_mort}
\end{subfigure}
\hfill
\begin{subfigure}[t]{0.49\textwidth}
    \centering
    \includegraphics[width=\linewidth]{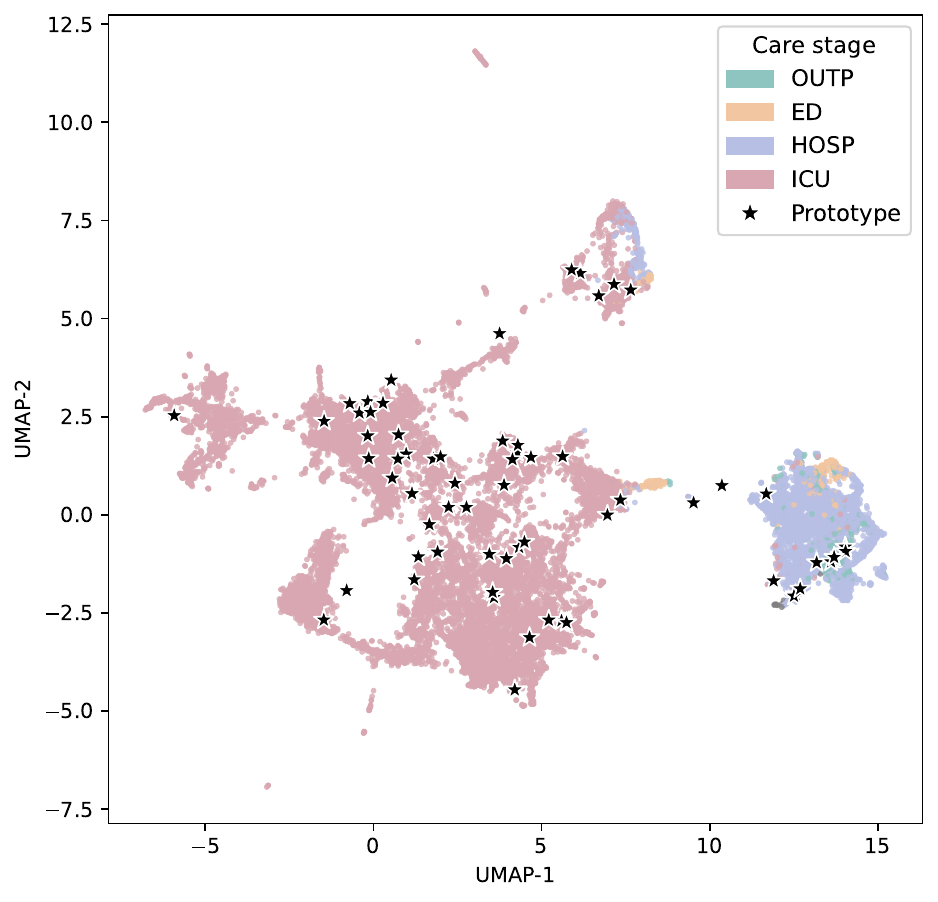}
    \caption{Long LOS ($\geq 7$ Days)}
    \label{fig:umap_los}
\end{subfigure}

\caption{
UMAP visualizations of history-chunk embeddings and learned prototypes across four clinical prediction tasks: 
(a) 1-year mortality post-discharge,
(b) ICU readmission within 30 days,
(c) in-hospital mortality, and
(d) long length of stay ($\geq 7$ days).
}
\label{fig:umap_chunking}
\end{figure}

Figure~\ref{fig:umap_chunking} visualizes the relationship between history chunk embeddings and learned prototypes across the four tasks. Overall, the embedding spaces exhibit partially separated care-stage structure, while the prototypes follow the geometry of the learned manifolds: they are concentrated in high-density regions, extend across distinct components, and rarely appear as isolated points. Their allocation is non-uniform and differs across tasks, ranging from broad coverage of the embedding space to greater concentration within particular regions. Prototypes also occur in both care-stage-dominant and mixed-stage regions, indicating that their organization is not reducible to care-stage labels alone. These projections provide qualitative evidence that the prototypes act as distributed reference points within the representation space, although UMAP geometry and prototype locations should not be interpreted as establishing the clinical meaning of individual prototypes.

\subsection{Qualitative Analysis of Prototype-Guided Retrieval}
\label{sec:qualitative}
This section provides a qualitative analysis of the prototype-guided retrieval mechanism in \texttt{EHR-RAGp}. To illustrate the behavior of the model, we present two representative case studies from the in-hospital mortality task, corresponding to positive and negative prediction examples. For each case, we analyze prototype usage, query-history alignment, chunk-level relevance weighting, and the contribution of visits and care stages to the final prediction. These examples provide insight into how \texttt{EHR-RAGp} organizes longitudinal patient history and integrates clinically relevant context during inference.

\subsubsection{Case of Positive Example}
\begin{figure}[!ht]
    \centering
    \includegraphics[width=\linewidth]{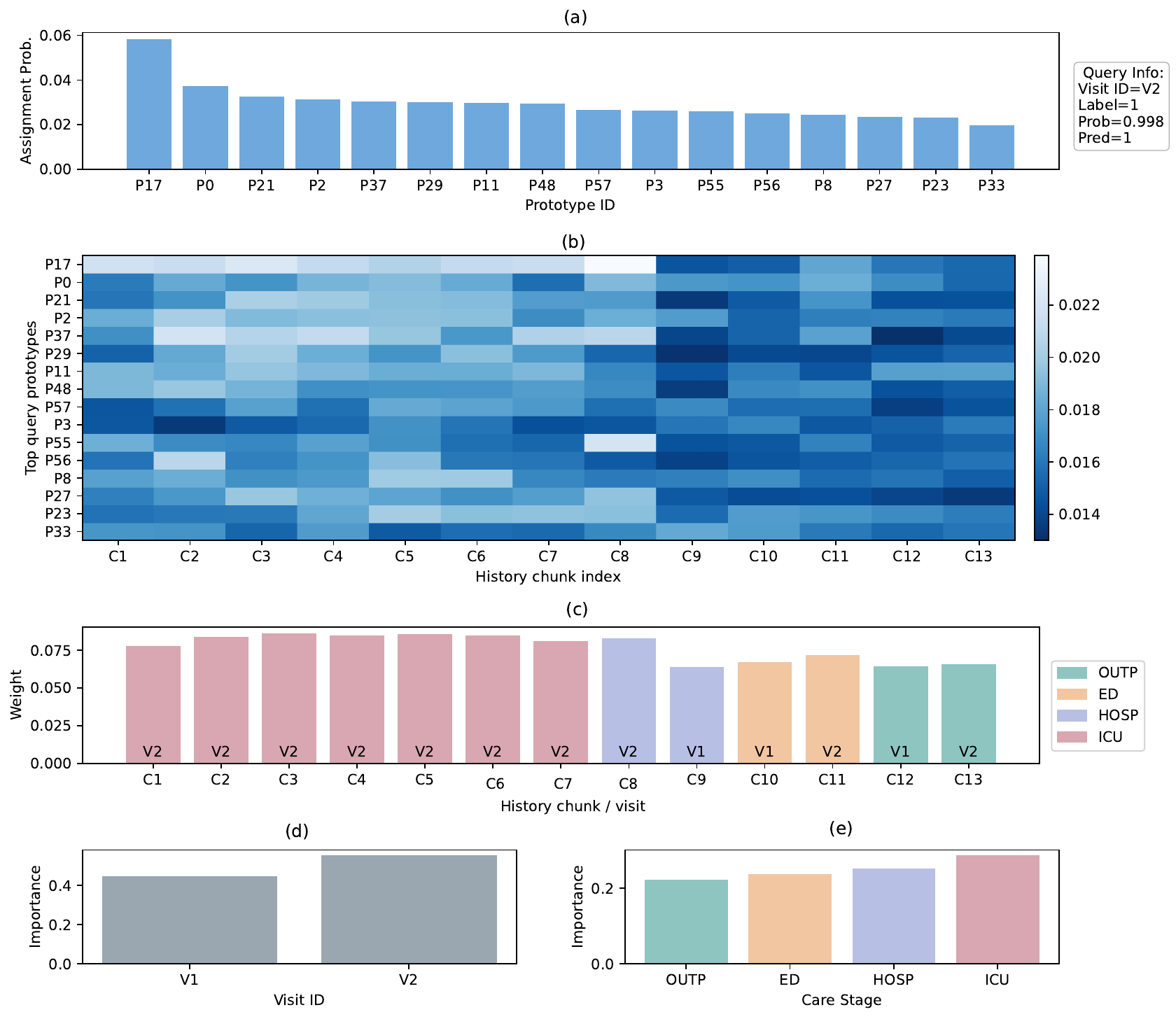}
    \caption{Qualitative analysis of prototype-guided retrieval for an in-hospital mortality positive example. (a) Top prototype assignments for the query segment. (b) Alignment between retrieved history chunks and the query prototypes. (c) Prototype-guided relevance weights assigned to retrieved chunks. (d) Aggregated contribution of retrieved visits. (e) Aggregated importance across care stages.}
    \label{fig:pos-exaple}
\end{figure}

Figure~\ref{fig:pos-exaple} presents a representative positive example from the in-hospital mortality task, corresponding to a correctly predicted positive case (\texttt{label=1}, \texttt{pred=1}) with high confidence (\texttt{prob=0.901}) for a query segment originating from Visit~V10 in the patient timeline. 

\subsubsection{Case of Negative Example}

\begin{figure}[!ht]
    \centering
    \includegraphics[width=\linewidth]{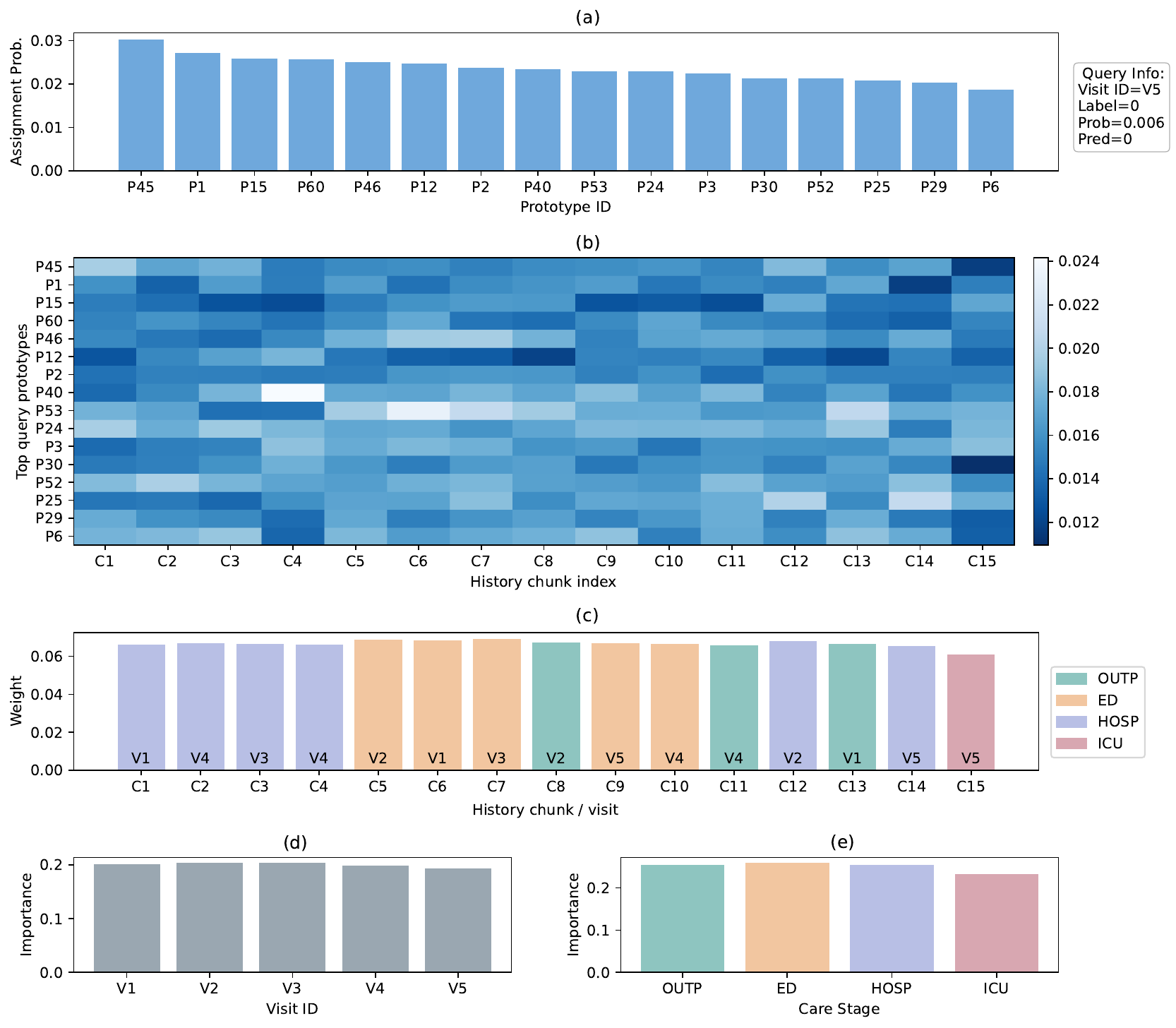}
    \caption{Qualitative analysis of prototype-guided retrieval for an in-hospital mortality negative example. (a) Top prototype assignments for the query segment. (b) Alignment between retrieved history chunks and the query prototypes. (c) Prototype-guided relevance weights assigned to retrieved chunks. (d) Aggregated contribution of retrieved visits. (e) Aggregated importance across care stages.}
    \label{fig:neg-example}
\end{figure}

Figure~\ref{fig:neg-example} presents a representative negative example from the in-hospital mortality task, corresponding to a correctly predicted negative case (\texttt{label=0}, \texttt{pred=0}) with low predicted probability (\texttt{prob=0.000}) for a query segment originating from Visit~V14 in the patient timeline. 

\subsection{Analysis of Prototype Regularization}
\label{sec:prot-reg-qual}
This section analyzes the effect of prototype regularization on the behavior of \texttt{EHR-RAGp}. Specifically, we compare the model with and without regularization from two perspectives: (1) global prototype utilization patterns, illustrating how regularization affects the distribution and diversity of prototype assignments, and (2) qualitative case studies, showing how regularization influences prototype-guided retrieval and chunk weighting for individual patient trajectories. Together, these analyses highlight the role of regularization in preventing prototype collapse and encouraging more balanced use of the latent prototype space.

\subsubsection{Global Prototypes Usage}

\begin{figure}[!t]
\centering
\begin{subfigure}[t]{0.49\textwidth}
    \centering
    \includegraphics[width=\linewidth,height=5cm]{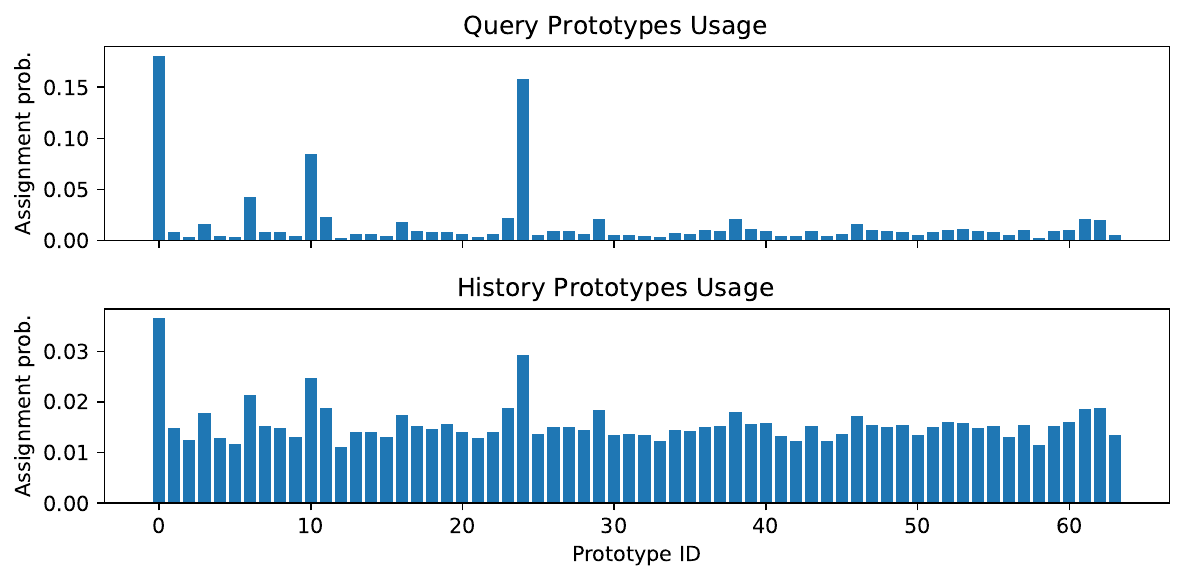}
    \caption{}
    \label{fig:proto-usage-without}
\end{subfigure}
\begin{subfigure}[t]{0.49\textwidth}
    \centering
    \includegraphics[width=\linewidth,height=5cm]{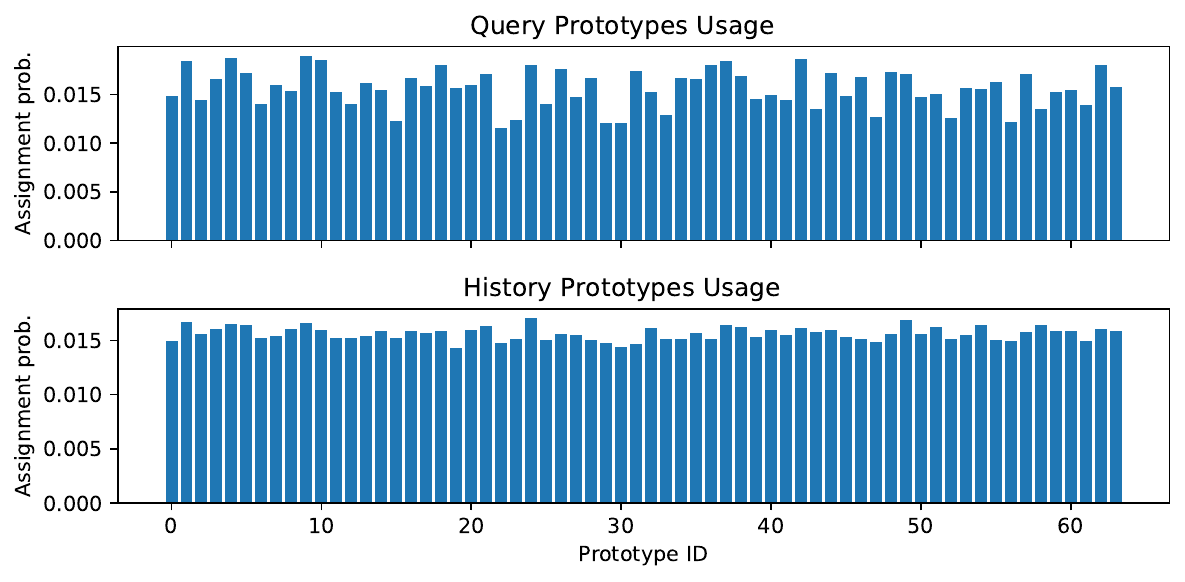}
    \caption{}
     \label{fig:proto-usage-with}
\end{subfigure}
\caption{Global prototype utilization patterns. (a) Without Regularization. (b) With regularization.} 
\label{fig:proto-usage}
\end{figure}

Figure \ref{fig:proto-usage} illustrates the effect of prototype regularization on global prototype utilization. Without regularization as in Figure \ref{fig:proto-usage-without}, query assignments exhibit highly imbalanced usage patterns, where a small subset of prototypes receive disproportionately large assignment probabilities while many prototypes remain minimally utilized. A similar, although less severe, imbalance is also observed for history chunk assignments. This behavior indicates partial prototype collapse, where the model concentrates representation learning around a limited number of latent prototypes. In turn, such a  situation makes agreement between history and query rather difficult due to  the sharper distribution of the query as compared to the history assignments. In contrast, enabling regularization, as in Figure \ref{fig:proto-usage-with}, produces a substantially more uniform assignment distribution across prototypes for both queries and history chunks. The resulting increase in prototype diversity suggests that regularization encourages broader utilization of the latent prototype space, leading to more balanced representation learning and improved retrieval behavior.

\subsubsection{Qualitative Analysis of Prototypes Usage}

\begin{figure}[!t]
    \centering
    \includegraphics[width=0.9\linewidth]{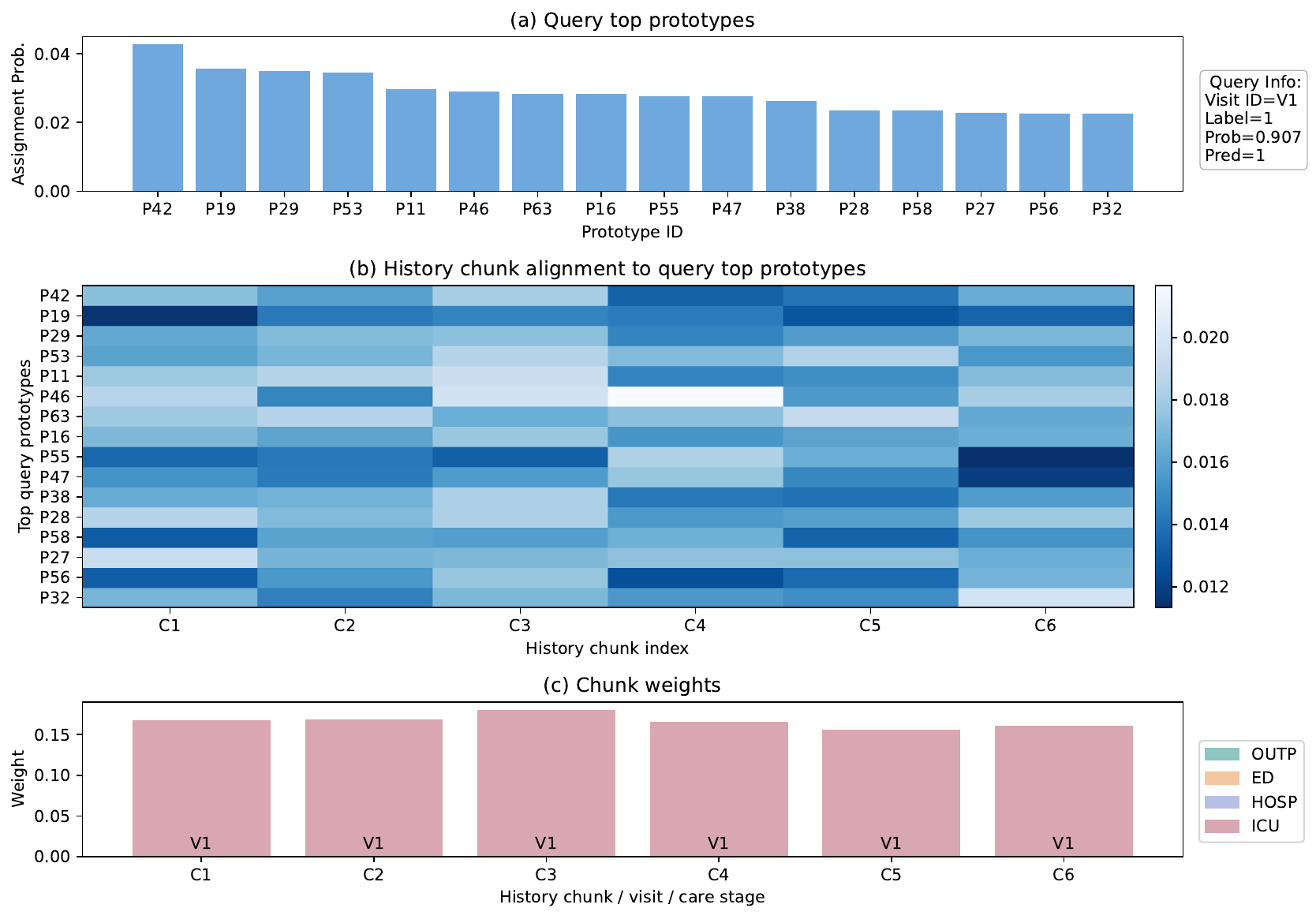}
    \caption{Qualitative analysis of sample prediction for long length of stay (7 days) task with regularization (a) Top prototype assignments for the query segment. (b) Alignment between retrieved history chunks and the query prototypes. (c) Prototype-guided relevance weights assigned to retrieved chunks.}
    \label{fig:los-reg-with}
\end{figure}

\begin{figure}[!t]
    \centering
    \includegraphics[width=0.9\linewidth]{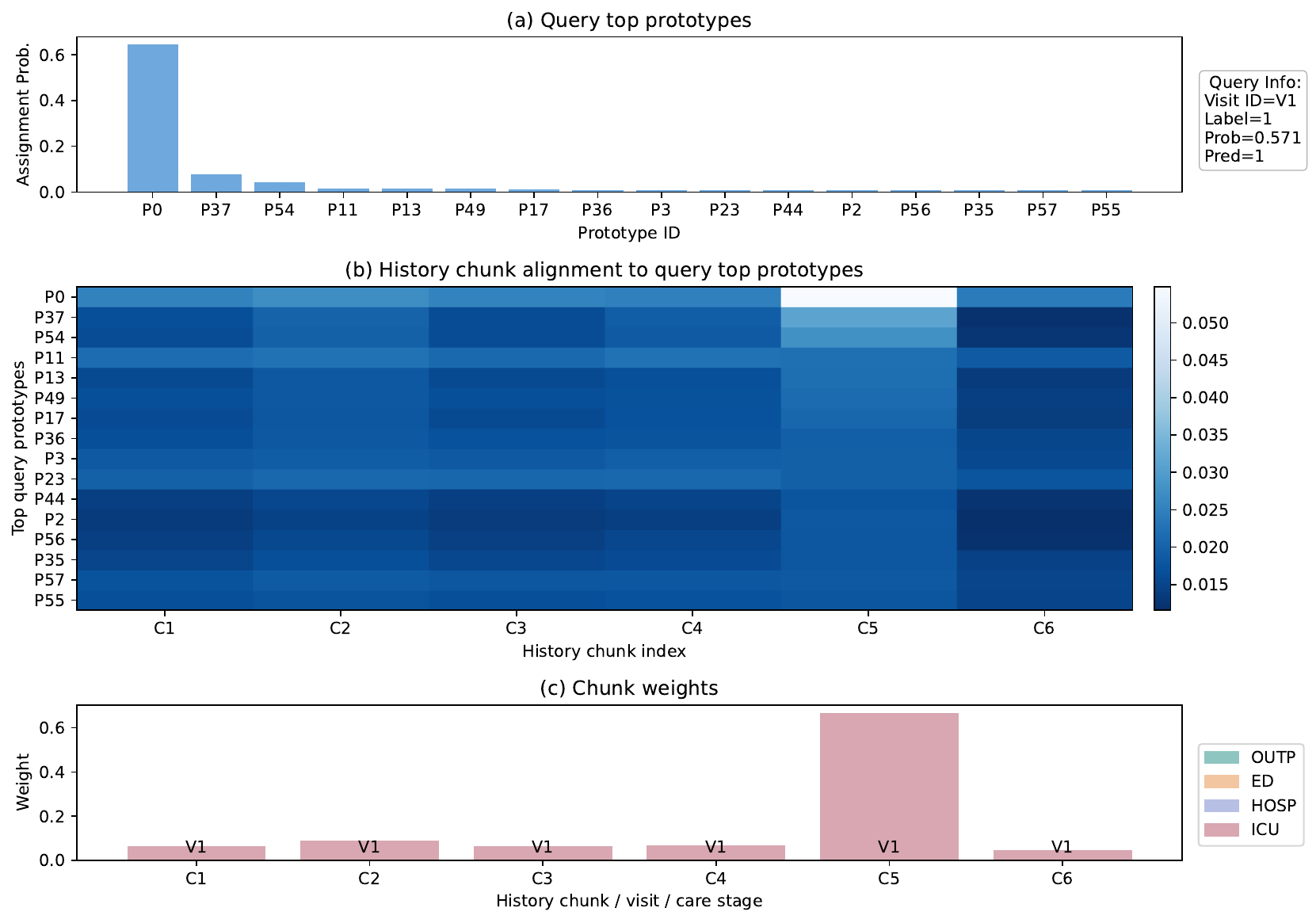}
    \caption{Qualitative analysis of sample prediction for long length of stay (7 days) task without regularization (a) Top prototype assignments for the query segment. (b) Alignment between retrieved history chunks and the query prototypes. (c) Prototype-guided relevance weights assigned to retrieved chunks.}
    \label{fig:los-reg-without}
\end{figure}

To further understand the role of usage regularization $\lambda_u$, we examine the model behavior with regularization, as in in Figure \ref{fig:los-reg-with},  and without regularization, as in Figure \ref{fig:los-reg-without}. In the settings where usage regularization is activated, prototype assignments are distributed across multiple latent prototypes, resulting in relatively balanced alignment scores and chunk weights across retrieved history segments. This behavior indicates that the model integrates information from several relevant historical chunks rather than relying on a single dominant prototype or retrieval path. On the other hand, without usage regularization, the query representation collapses strongly toward a single prototype (P0), which leads to highly concentrated retrieval behavior where few retrieved chunk receives the majority of the final weight while the remaining chunks contribute minimally. With respect to the model performance, it can be observed that despite the collapse the model can still provide correct prediction but less confident. Overall, these examples qualitatively demonstrate how prototype regularization stabilizes retrieval dynamics, encourages broader utilization of patient history, and prevents highly concentrated dependence on a narrow subset of latent prototypes, which in turn deteriorates the retrieval quality.

\newpage
\subsection{EHRSHOT Benchmark Results}

 Comparisons in Table \ref{tab:ehrshot_sig} are descriptive and based on the reported AUROC point estimates. Although all methods are evaluated using the predefined EHRSHOT splits, we do not claim that differences between \texttt{EHR-RAGp} and the long-context baselines are statistically significant, as paired predictions for these baselines are unavailable. The confidence intervals and significance indicators reported for the long-context models reflect their original comparisons with CLMBR-\textit{t}-base and are included only for completeness.

\subsection{Additional Results  on MIMIC-IV Dataset}
Table \ref{tab:main_results2} reports the bootstrapped confidence intervals computed for all baselines models reported in Table \ref{tab:mimic_results}, whereas Table \ref{tab:w/wo1} reports the confidence interval for all models reported in Table \ref{tab:w/wo}. Table \ref{tab:chunking} complements the results presented in table \ref{tab:chunking1} and report the confidence intervals for each reported result. Table \ref{tab:main_results1} report the p-values computed for all baselines used in our experiments.

\subsection{Additional ablations on MIMIC-IV Dataset}

Table \ref{tab:mechanism} complements the results provided in Table \ref{tab:retrieval_mech} by comparing prototype effect for ICU-Readmit 30d and In-hospital Mortality tasks. Table \ref{tab:num_proto} presents ablation with number of prototypes, while Table \ref{tab:softmax} shows the effect of softmax temperature on the performance.








\begin{table}[t]
\centering
\scriptsize
\setlength{\tabcolsep}{3.pt}
\renewcommand{\arraystretch}{0.82}
\caption{
AUROC differences relative to CLMBR-\textit{t}-base on operational-outcome
and new-diagnosis assignment tasks. For each long-context baseline, we
report the maximum context length evaluated by Wornow et
al.~\citep{wornow2024context}. The best and second-best results for each
task are shown in \textbf{bold} and \underline{underlined}, respectively.
A checkmark indicates that the reported 95\% confidence interval excludes zero.
}
\label{tab:ehrshot_max_context}
\resizebox{0.9\linewidth}{!}{
\begin{tabular}{cllccc}
\toprule
\textbf{Context}
& \textbf{Model}
& \textbf{Task}
& \shortstack{\textbf{$\Delta$ over}\\
\textbf{CLMBR-\textit{t}-base}}
& \textbf{95\% CI}
& \textbf{Sig.} \\
\midrule

\multirow{9}{*}{4096}
& \multirow{9}{*}{LLaMA}
& ICU Admission      & -0.003 & (-0.026, 0.021) & \\
& & Long LOS           & -0.004 & (-0.018, 0.010) & \\
& & 30-day Readmission & \textbf{0.013} & (0.002, 0.026) & $\checkmark$ \\
\addlinespace[1pt]
& & Acute MI           & 0.015 & (-0.008, 0.038) & \\
& & Celiac             & \underline{0.226} & (0.097, 0.365) & $\checkmark$ \\
& & Hyperlipidemia     & 0.016 & (-0.002, 0.036) & \\
& & Hypertension       & \underline{0.004} & (-0.013, 0.021) & \\
& & Lupus              & -0.023 & (-0.097, 0.049) & \\
& & Pancreatic Cancer  & -0.008 & (-0.056, 0.033) & \\

\midrule

\multirow{9}{*}{4096}
& \multirow{9}{*}{GPT}
& ICU Admission      & \underline{0.011} & (-0.021, 0.044) & \\
& & Long LOS           & -0.001 & (-0.014, 0.014) & \\
& & 30-day Readmission & 0.004 & (-0.009, 0.015) & \\
\addlinespace[1pt]
& & Acute MI           & 0.006 & (-0.022, 0.033) & \\
& & Celiac             & 0.149 & (0.041, 0.278) & $\checkmark$ \\
& & Hyperlipidemia     & 0.012 & (-0.018, 0.043) & \\
& & Hypertension       & \underline{0.004} & (-0.015, 0.024) & \\
& & Lupus              & -0.008 & (-0.095, 0.088) & \\
& & Pancreatic Cancer  & 0.027 & (-0.008, 0.062) & \\

\midrule

\multirow{9}{*}{16384}
& \multirow{9}{*}{Hyena}
& ICU Admission      & -0.110 & (-0.147, -0.075) & $\checkmark$ \\
& & Long LOS           & -0.048 & (-0.068, -0.029) & $\checkmark$ \\
& & 30-day Readmission & -0.048 & (-0.067, -0.026) & $\checkmark$ \\
\addlinespace[1pt]
& & Acute MI           & -0.100 & (-0.145, -0.053) & $\checkmark$ \\
& & Celiac             & 0.176 & (0.029, 0.318) & $\checkmark$ \\
& & Hyperlipidemia     & -0.016 & (-0.069, 0.034) & \\
& & Hypertension       & -0.071 & (-0.125, -0.023) & $\checkmark$ \\
& & Lupus              & -0.145 & (-0.268, -0.017) & $\checkmark$ \\
& & Pancreatic Cancer  & -0.073 & (-0.148, 0.006) & \\

\midrule

\multirow{9}{*}{16384}
& \multirow{9}{*}{Mamba}
& ICU Admission      & 0.007 & (-0.028, 0.040) & \\
& & Long LOS           & \underline{0.013} & (-0.005, 0.029) & \\
& & 30-day Readmission & 0.005 & (-0.008, 0.017) & \\
\addlinespace[1pt]
& & Acute MI           & \underline{0.016} & (-0.005, 0.036) & \\
& & Celiac             & 0.194 & (0.108, 0.333) & $\checkmark$ \\
& & Hyperlipidemia     & \underline{0.023} & (-0.013, 0.058) & \\
& & Hypertension       & 0.003 & (-0.018, 0.023) & \\
& & Lupus              & \underline{0.037} & (-0.056, 0.132) & \\
& & Pancreatic Cancer  & \underline{0.053} & (0.024, 0.087) & $\checkmark$ \\

\midrule

\multirow{9}{*}{$512+ (24 \times 256)$}
& \multirow{9}{*}{\texttt{EHR-RAGp}}
& ICU Admission      & \textbf{0.033} & -- & -- \\
& & Long LOS           & \textbf{0.031} & -- & -- \\
& & 30-day Readmission & \underline{0.006} & -- & -- \\
\addlinespace[1pt]
& & Acute MI           & \textbf{0.032} & -- & -- \\
& & Celiac             & \textbf{0.228} & -- & -- \\
& & Hyperlipidemia     & \textbf{0.046} & -- & -- \\
& & Hypertension       & \textbf{0.031} & -- & -- \\
& & Lupus              & \textbf{0.116} & -- & -- \\
& & Pancreatic Cancer  & \textbf{0.060} & -- & -- \\

\bottomrule
\end{tabular}
}
\label{tab:ehrshot_sig}
\end{table}

\begin{table*}[t!]
\centering
\caption{Performance comparison between \texttt{EHR-RAGp} and all baselines. Best results are marked in \textbf{bold}. Second best results are \underline{underlined}. }
\resizebox{\linewidth}{!}{
\begin{tabular}{lcccccccc}
\toprule
\multirow{1}{*}{\textbf{Model}}  &\multicolumn{2}{c}{\textbf{ICU-Readmit 30d}} 
& \multicolumn{2}{c}{\textbf{In-hospital Mortality}} 
& \multicolumn{2}{c}{\textbf{Long LOS 7d}} 
& \multicolumn{2}{c}{\textbf{1YR Mortality}} \\

 & \textbf{AUROC \scriptsize(CI)} & \textbf{AUPRC \scriptsize(CI)} 
& \textbf{AUROC \scriptsize(CI)} & \textbf{AUPRC \scriptsize(CI)} 
& \textbf{AUROC \scriptsize(CI)} & \textbf{AUPRC \scriptsize(CI)} 
 & \textbf{AUROC \scriptsize(CI)} & \textbf{AUPRC \scriptsize(CI)} \\
\midrule

DescEmb I  
& 0.661 \scriptsize(0.630, 0.690) &  0.078 \scriptsize(0.064, 0.095)
& 0.904 \scriptsize(0.890, 0.915) &  0.618 \scriptsize(0.578, 0.656)
& 0.822 \scriptsize(0.806, 0.836) &  0.405 \scriptsize(0.369, 0.440)
& 0.754 \scriptsize(0.738, 0.770) & 0.376 \scriptsize(0.347, 0.406)\\


DescEmb II 
& 0.606 \scriptsize(0.573, 0.638) &  0.071 \scriptsize(0.058, 0.091)
& 0.920 \scriptsize(0.909, 0.931) &  0.648 \scriptsize(0.607, 0.686)
& 0.838 \scriptsize(0.824, 0.852) &  0.441 \scriptsize(0.405, 0.480)
& 0.805 \scriptsize(0.791, 0.820) &  0.471 \scriptsize(0.440, 0.504) \\

GenHPF   
& 0.697 \scriptsize(0.667, 0.729) &  \underline{0.128 \scriptsize(0.099, 0.169)}
& 0.915 \scriptsize(0.904, 0.926) &  0.641 \scriptsize(0.604, 0.681)
& 0.843 \scriptsize(0.828, 0.856) &  0.456 \scriptsize(0.419, 0.494)
& 0.781 \scriptsize(0.767, 0.795) &  0.417 \scriptsize(0.388, 0.448)\\

REMed 
& 0.561 \scriptsize(0.528, 0.595) &  0.060 \scriptsize(0.049, 0.079)
& 0.865 \scriptsize(0.851, 0.880) &  0.501 \scriptsize(0.462, 0.542)
& 0.786 \scriptsize(0.770, 0.803) &  0.348 \scriptsize(0.317, 0.383)
& 0.680 \scriptsize(0.662, 0.697) &  0.280 \scriptsize(0.257, 0.303)\\

Med-BERT 
& 0.664 \scriptsize(0.633, 0.695) &  0.081 \scriptsize(0.067, 0.105)
& 0.893 \scriptsize(0.879, 0.907) &  0.579 \scriptsize(0.540, 0.621)
& 0.813 \scriptsize(0.798, 0.829) &  0.412 \scriptsize(0.378, 0.446)
& 0.729 \scriptsize(0.714, 0.745) &  0.338 \scriptsize(0.312, 0.366)\\

CEHR-BERT   
& 0.604 \scriptsize(0.573, 0.637) &  0.061 \scriptsize(0.051, 0.075)
& \underline{0.929 \scriptsize(0.919, 0.938)} &  \underline{0.676 \scriptsize(0.641, 0.714)}
& 0.840 \scriptsize(0.825, 0.854) &  0.435 \scriptsize(0.396, 0.473)
& \underline{0.806 \scriptsize(0.791, 0.820)} &  \underline{0.473 \scriptsize(0.440, 0.506)}\\

BEHRT   
& 0.668 \scriptsize(0.638, 0.698) &  0.077 \scriptsize(0.065, 0.094)
& 0.902 \scriptsize(0.889, 0.913) &  0.609 \scriptsize(0.569, 0.650)
& 0.824 \scriptsize(0.809, 0.840) &  0.444 \scriptsize(0.408, 0.481)
& 0.767 \scriptsize(0.751, 0.782) &  0.382 \scriptsize(0.353, 0.413)\\

Hi-BEHRT   
& 0.620 \scriptsize(0.587, 0.652) &  0.069 \scriptsize(0.058, 0.087)
& 0.886 \scriptsize(0.873, 0.899) &  0.542 \scriptsize(0.501, 0.585)
& 0.798 \scriptsize(0.780, 0.816) &  0.392 \scriptsize(0.356, 0.430)
& 0.703 \scriptsize(0.687, 0.721) &  0.298 \scriptsize(0.275, 0.325) \\


EHRMamba  
& 0.665 \scriptsize(0.633, 0.695) &  0.082 \scriptsize(0.067, 0.100)
& \underline{0.929 \scriptsize(0.918, 0.938)} &  0.650 \scriptsize(0.611, 0.690)
& 0.829 \scriptsize(0.813, 0.844) &  0.424 \scriptsize(0.385, 0.464)

& 0.796 \scriptsize(0.782, 0.809) & 0.429 \scriptsize(0.400, 0.462) \\

ModernBERT  
& 0.604 \scriptsize(0.571, 0.635) &  0.068 \scriptsize(0.056, 0.087)
& 0.920 \scriptsize(0.910, 0.930) &  0.640 \scriptsize(0.604, 0.679)
& 0.827 \scriptsize(0.811, 0.842) &  0.440 \scriptsize(0.404, 0.475)
& 0.765 \scriptsize(0.750, 0.780) & 0.381 \scriptsize(0.354, 0.414)\\


LongFormer  
& 0.645 \scriptsize(0.613, 0.679) &  0.095 \scriptsize(0.074, 0.129)
& 0.910 \scriptsize(0.900, 0.921) &  0.577 \scriptsize(0.536, 0.620)
& 0.821 \scriptsize(0.806, 0.837) &  0.423 \scriptsize(0.385, 0.463) 
& 0.760 \scriptsize(0.744, 0.775) &  0.377 \scriptsize(0.349, 0.409)\\

BigBird 
& 0.719 \scriptsize(0.691, 0.746) &  0.118 \scriptsize(0.094, 0.152)
& 0.913 \scriptsize(0.903, 0.924) &  0.643 \scriptsize(0.606, 0.680)
& 0.830 \scriptsize(0.813, 0.844) &  0.444 \scriptsize(0.407, 0.481)
& 0.787 \scriptsize(0.773, 0.803) & 0.424 \scriptsize(0.393, 0.455)\\

RoFormer 
& 0.699 \scriptsize(0.670, 0.728) &  0.114 \scriptsize(0.088, 0.149)
& 0.927 \scriptsize(0.917, 0.937) &  0.672 \scriptsize(0.637, 0.706)
& \underline{0.846 \scriptsize(0.830, 0.859)} &  0.461 \scriptsize(0.424, 0.502)
& \underline{0.806 \scriptsize(0.791, 0.820)} &  0.460 \scriptsize(0.427, 0.495)\\



RoBERTa 
& \underline{0.721 \scriptsize(0.690, 0.750)} &  0.125 \scriptsize(0.099, 0.162)
& 0.925 \scriptsize(0.916, 0.934) &  0.671 \scriptsize(0.637, 0.705)
& 0.841 \scriptsize(0.827, 0.856) &  \underline{0.465 \scriptsize(0.430, 0.506)}
& 0.805 \scriptsize(0.791, 0.818) &  0.466 \scriptsize(0.436, 0.498)\\

\midrule

\texttt{EHR-RAGp} (Ours) 
& \textbf{0.750 \scriptsize(0.723, 0.774)} & \textbf{0.141 \scriptsize(0.115, 0.184)}
& \textbf{0.950 \scriptsize(0.941, 0.956)} &  \textbf{0.736 \scriptsize(0.701, 0.771)} 
& \textbf{0.867 \scriptsize(0.853, 0.878)} & \textbf{0.509 \scriptsize(0.472, 0.545)}
& \textbf{0.853 \scriptsize(0.841, 0.864)} & \textbf{0.564 \scriptsize(0.531, 0.595)} \\
\bottomrule
\end{tabular}}
\label{tab:main_results2}
\end{table*}

\begin{table*}[!t]
\centering
\caption{Performance gains of existing EHR baseline with \texttt{EHR-RAGp}.}
\resizebox{\linewidth}{!}{
\begin{tabular}{lccccccccc}
\toprule
\multirow{2}{*}{\textbf{Model}} 
& \multirow{2}{*}{\textbf{\texttt{EHR-RAGp}}}  & \multicolumn{2}{c}{\textbf{ICU-Readmit 30d}} 
& \multicolumn{2}{c}{\textbf{In-hospital Mortality}} 
& \multicolumn{2}{c}{\textbf{Long LOS 7d}} 
& \multicolumn{2}{c}{\textbf{1YR Mortality}} \\

& & \textbf{AUROC \scriptsize(CI)} & \textbf{AUPRC \scriptsize(CI)} 
& \textbf{AUROC \scriptsize(CI)} & \textbf{AUPRC \scriptsize(CI)} 
& \textbf{AUROC \scriptsize(CI)} & \textbf{AUPRC \scriptsize(CI)} 
 & \textbf{AUROC \scriptsize(CI)} & \textbf{AUPRC \scriptsize(CI)} \\
\midrule

\multirow{2}{*}{Med-BERT} 
& $\times$ 
& 0.664 \scriptsize(0.633, 0.695) &  0.081 \scriptsize(0.067, 0.105)
& 0.893 \scriptsize(0.879, 0.907) &  0.579 \scriptsize(0.540, 0.621)
& 0.813 \scriptsize(0.798, 0.829) &  0.412 \scriptsize(0.378, 0.446)
& 0.729 \scriptsize(0.714, 0.745) &  0.338 \scriptsize(0.312, 0.366)\\

& $\surd$  
& \textbf{0.715 \scriptsize(0.687, 0.742)} & \textbf{0.117 \scriptsize(0.093, 0.152)}
& \textbf{0.938 \scriptsize(0.929, 0.947)} &  \textbf{0.702 \scriptsize(0.663, 0.736)}
& \textbf{0.843 \scriptsize(0.829, 0.858)} &  \textbf{0.455 \scriptsize(0.421, 0.494)}
& \textbf{0.823 \scriptsize(0.810, 0.836)} & \textbf{0.476 \scriptsize(0.445, 0.510)} \\

\midrule
\multirow{2}{*}{CEHR-BERT} 
& $\times$ 
& 0.604 \scriptsize(0.573, 0.637) &  0.061 \scriptsize(0.051, 0.075)
& 0.929 \scriptsize(0.919, 0.938) &  0.676 \scriptsize(0.641, 0.714)
& 0.840 \scriptsize(0.825, 0.854) &  0.435 \scriptsize(0.396, 0.473)
& 0.806 \scriptsize(0.791, 0.820) &  0.473 \scriptsize(0.440, 0.506)\\

& $\surd$  
& \textbf{0.698 \scriptsize(0.669, 0.727)} & \textbf{0.102 \scriptsize(0.082, 0.130)}
& \textbf{0.941 \scriptsize(0.932, 0.949)} &  \textbf{0.715 \scriptsize(0.676, 0.751)}
& \textbf{0.848 \scriptsize(0.835, 0.862)} &  \textbf{0.452 \scriptsize(0.417, 0.491)}
& \textbf{0.821 \scriptsize(0.807, 0.834)} & \textbf{0.495 \scriptsize(0.465, 0.527)} \\

\midrule
\multirow{2}{*}{BEHRT} 
& $\times$ 
& 0.668 \scriptsize(0.638, 0.698) &  0.077 \scriptsize(0.065, 0.094)
& 0.902 \scriptsize(0.889, 0.913) &  0.609 \scriptsize(0.569, 0.650)
& 0.824 \scriptsize(0.809, 0.840) &  0.444 \scriptsize(0.408, 0.481)
& 0.767 \scriptsize(0.751, 0.782) &  0.382 \scriptsize(0.353, 0.413)\\

& $\surd$  
& \textbf{0.723 \scriptsize(0.692, 0.751)} & \textbf{0.126 \scriptsize(0.098, 0.161)}
& \textbf{0.941 \scriptsize(0.933, 0.949)} &  \textbf{0.711 \scriptsize(0.677, 0.748)}
& \textbf{0.849 \scriptsize(0.836, 0.863)} &  \textbf{0.465 \scriptsize(0.427, 0.503)}
& \textbf{0.826 \scriptsize(0.813, 0.838)} &  \textbf{0.490 \scriptsize(0.458, 0.524)}\\
\midrule

\multirow{2}{*}{Hi-BEHRT} 
& $\times$ 
& 0.620 \scriptsize(0.587, 0.652) &  0.069 \scriptsize(0.058, 0.087)
& 0.886 \scriptsize(0.873, 0.899) &  0.542 \scriptsize(0.501, 0.585)
& 0.798 \scriptsize(0.780, 0.816) &  0.392 \scriptsize(0.356, 0.430)
& 0.703 \scriptsize(0.687, 0.721) &  0.298 \scriptsize(0.275, 0.325) \\

& $\surd$  
& \textbf{0.684 \scriptsize(0.652, 0.716)} & \textbf{0.100 \scriptsize(0.080, 0.128)}
& \textbf{0.934 \scriptsize(0.924, 0.942)} &  \textbf{0.691 \scriptsize(0.655, 0.728)}
& \textbf{0.835 \scriptsize(0.821, 0.848)} &  \textbf{0.443 \scriptsize(0.408, 0.478)}
& \textbf{0.786 \scriptsize(0.772, 0.800)} & \textbf{0.401 \scriptsize(0.37, 0.435)} \\

\bottomrule

\end{tabular}
}

\label{tab:w/wo1}
\end{table*}

\begin{table*}[t!]
\centering
\caption{{Impact of choice of chunking strategy on model performance across all tasks.} }
\resizebox{0.99\linewidth}{!}{
\begin{tabular}{lcccccccc}
\toprule
\multirow{1}{*}{\textbf{Model}} 
& \multicolumn{2}{c}{\textbf{ICU-Readmit 30d}} 
& \multicolumn{2}{c}{\textbf{In-hospital Mortality}} 
& \multicolumn{2}{c}{\textbf{Long LOS 7d}} 
& \multicolumn{2}{c}{\textbf{1YR Mortality}} \\

& \textbf{AUROC \scriptsize(CI)} & \textbf{AUPRC \scriptsize(CI)} 
& \textbf{AUROC \scriptsize(CI)} & \textbf{AUPRC \scriptsize(CI)} 
& \textbf{AUROC \scriptsize(CI)} & \textbf{AUPRC \scriptsize(CI)} 
 & \textbf{AUROC \scriptsize(CI)} & \textbf{AUPRC \scriptsize(CI)} \\
\midrule

Event-Based  
& \textbf{0.750 \scriptsize(0.723, 0.774)} & \textbf{0.141 \scriptsize(0.115, 0.184)}
& \textbf{0.950 \scriptsize(0.941, 0.956)} &  0.736 \scriptsize(0.701, 0.771)
& \textbf{0.867 \scriptsize(0.854, 0.878)} &  0.506 \scriptsize(0.470, 0.542)
& 0.851 \scriptsize(0.838, 0.864) & 0.553 \scriptsize(0.521, 0.588)\\

Time-Based   
& 0.741 \scriptsize(0.712, 0.767) & 0.129 \scriptsize(0.103, 0.163)
& 0.946 \scriptsize(0.938, 0.954) &  0.738 \scriptsize(0.701, 0.766)
& 0.862 \scriptsize(0.848, 0.875) & 0.507 \scriptsize(0.469, 0.541)
& 0.847 \scriptsize(0.834, 0.859) & 0.538 \scriptsize(0.506, 0.573) \\

Visit-Level   
& 0.748 \scriptsize(0.720, 0.778) & 0.140 \scriptsize(0.110, 0.181)
& 0.948 \scriptsize(0.940, 0.955) & 0.738 \scriptsize(0.704, 0.772)
& \textbf{0.867 \scriptsize(0.853, 0.878)} & \textbf{0.509 \scriptsize(0.472, 0.545)}
& 0.852 \scriptsize(0.839, 0.864) & 0.553 \scriptsize(0.522, 0.583) \\
 
Care-Stage   
& 0.747 \scriptsize(0.719, 0.774) & 0.138 \scriptsize(0.111, 0.177)
& 0.947 \scriptsize(0.938, 0.955) & \textbf{0.739 \scriptsize(0.704, 0.771)}
& 0.865 \scriptsize(0.851, 0.878) & 0.506 \scriptsize(0.470, 0.544)
& \textbf{0.853 \scriptsize(0.841, 0.864)} & \textbf{0.564 \scriptsize(0.531, 0.595)}  \\
\bottomrule
\end{tabular}}
\label{tab:chunking}
\end{table*}

\begin{table*}[t!]
\centering
\caption{Paired-bootstrap significance of the performance differences between \texttt{EHR-RAGp} and each baseline.}
\resizebox{\linewidth}{!}{
\begin{tabular}{lcccccccc}
\toprule
\textbf{Model}
& \multicolumn{2}{c}{\textbf{ICU-Readmit 30d}}
& \multicolumn{2}{c}{\textbf{In-hospital Mortality}}
& \multicolumn{2}{c}{\textbf{Long LOS 7d}}
& \multicolumn{2}{c}{\textbf{1YR Mortality}} \\

& \textbf{AUROC} & \textbf{AUPRC}
& \textbf{AUROC} & \textbf{AUPRC}
& \textbf{AUROC} & \textbf{AUPRC}
& \textbf{AUROC} & \textbf{AUPRC} \\
\midrule

DescEmb I
&  $<0.0001$ &  $0.0014$
&  $<0.0001$ &  $<0.0001$
&  $<0.0001$ &  $<0.0001$
&  $<0.0001$ &  $<0.0001$ \\

DescEmb II
&  $<0.0001$ &  $0.0002$
&  $<0.0001$ &  $<0.0001$
&  $<0.0001$ &  $<0.0001$
&  $<0.0001$ &  $<0.0001$ \\

GenHPF 
&  $0.0004$ &  $0.2395$
&  $<0.0001$ &  $<0.0001$
&  $<0.0001$ &  $<0.0001$
&  $<0.0001$ &  $<0.0001$ \\

REMed 
&  $<0.0001$ &  $0.0004$
&  $<0.0001$ &  $<0.0001$
&  $<0.0001$ &  $<0.0001$
&  $<0.0001$ &  $<0.0001$ \\

Med-BERT 
&  $<0.0001$ &  $0.0006$
&  $<0.0001$ &  $<0.0001$
&  $<0.0001$ &  $<0.0001$
&  $<0.0001$ &  $<0.0001$ \\

CEHR-BERT 
&  $<0.0001$ &  $0.0006$
&  $<0.0001$ &  $<0.0001$
&  $<0.0001$ &  $<0.0001$
&  $<0.0001$ &  $<0.0001$ \\

BEHRT 
&  $<0.0001$ &  $0.0011$
&  $<0.0001$ &  $<0.0001$
&  $<0.0001$ &  $<0.0001$
&  $<0.0001$ &  $<0.0001$ \\

Hi-BEHRT 
& $<0.0001$ & $0.0018$
& $<0.0001$ & $<0.0001$
& $<0.0001$ & $<0.0001$
& $<0.0001$ & $<0.0001$ \\

EHRMamba 
& $<0.0001$ & $0.0016$
& $<0.0001$ & $<0.0001$
& $<0.0001$ & $<0.0001$
& $<0.0001$ & $<0.0001$ \\

ModernBERT 
& $<0.0001$ & $0.0008$
& $<0.0001$ & $<0.0001$
& $<0.0001$ & $<0.0001$
& $<0.0001$ & $<0.0001$ \\

Longformer 
& $<0.0001$ & $0.0009$
& $<0.0001$ & $<0.0001$
& $<0.0001$ & $<0.0001$
&  $<0.0001$ & $<0.0001$ \\

BigBird 
&  $0.0098$ & $0.0355$
&  $<0.0001$ & $<0.0001$
&  $<0.0001$ & $<0.0001$
&  $<0.0001$ & $<0.0001$ \\

RoFormer 
&  $<0.0001$ &  $0.0163$
&  $<0.0001$ &  $<0.0001$
&  $<0.0001$ &  $<0.0001$
&  $<0.0001$ &  $<0.0001$ \\

RoBERTa 
& $0.0013$ & $0.0246$
& $<0.0001$ & $<0.0001$
& $<0.0001$ & $<0.0001$
& $<0.0001$ &  $<0.0001$ \\

\bottomrule
\end{tabular}
}

\parbox{\linewidth}{\footnotesize
}

\label{tab:main_results1}
\end{table*}

\begin{table}[t]
\centering
\caption{Ablation of retrieval mechanisms. $^\dagger$denotes p$<$ 0.05 compared to \texttt{EHR-RAGp}.}
\label{tab:retrieval_mech1}
\begin{tabular}{lcccc}
\toprule
\multirow{2}{*}{\textbf{Configuration}}
& \multicolumn{2}{c}{\textbf{ICU Readmission 30d}}
& \multicolumn{2}{c}{\textbf{1-Year Mortality}} \\
& \textbf{AUROC}
& \textbf{AUPRC}
& \textbf{AUROC}
& \textbf{AUPRC} \\
\midrule
w/o Retrieval
& 0.721$^\dagger$
& 0.125$^\dagger$
& 0.925$^\dagger$
& 0.671$^\dagger$ \\

w/o Prototypes
& 0.735$^\dagger$
& 0.134
& 0.941$^\dagger$
& 0.713$^\dagger$ \\

w/ Prototypes
& 0.750
& 0.141
& 0.950
& 0.736 \\
\bottomrule
\end{tabular}
\label{tab:mechanism}
\end{table}

\begin{table}[t]
\centering
\caption{Ablation of the number of prototypes.}
\label{tab:num-proto}
\begin{tabular}{ccccc}
\toprule
\multirow{2}{*}{\textbf{\# Prototypes}}
& \multicolumn{2}{c}{\textbf{ICU Readmission 30d}}
& \multicolumn{2}{c}{\textbf{1-Year Mortality}} \\
& \textbf{AUROC}
& \textbf{AUPRC}
& \textbf{AUROC}
& \textbf{AUPRC} \\
\midrule
64
& 0.744
& \textbf{0.139}
& 0.850
& 0.551 \\

128
& \textbf{0.745}
& 0.137
& 0.844
& 0.536 \\

256
& 0.742
& 0.137
& 0.845
& 0.536 \\

512
& 0.727
& 0.129
& \textbf{0.851}
& \textbf{0.555} \\
\bottomrule
\end{tabular}
\label{tab:num_proto}
\end{table}

\begin{table}[t]
\centering
\caption{Effect of the softmax temperature on model performance.}
\label{tab:softmax_temp}
\begin{tabular}{ccccc}
\toprule
\multirow{2}{*}{\textbf{$T_s$}}
& \multicolumn{2}{c}{\textbf{1-Year Mortality}}
& \multicolumn{2}{c}{\textbf{In-Hospital Mortality}} \\
& \textbf{AUROC}
& \textbf{AUPRC}
& \textbf{AUROC}
& \textbf{AUPRC} \\
\midrule
0.10
& 0.845 & 0.540
& 0.947 & 0.728 \\

0.15
& 0.852 & 0.552
& 0.948 & 0.730 \\

0.20
& \textbf{0.853} & \textbf{0.564}
& 0.948 & 0.732 \\

0.25
& 0.851 & 0.553
& \textbf{0.950} & \textbf{0.736} \\
\bottomrule
\end{tabular}
\label{tab:softmax}
\end{table}

\end{document}